# Predicting Credit Spreads and Ratings with Machine Learning: The Role of Non-Financial Data

Yanran WU[1], Xinlei ZHANG[1], Quanyi XU[1], Qianxin YANG[1], Chao ZHANG[2*]

1. Business School, Beijing Normal University, No.19, Xinjiekouwai St, Haidian District, Beijing, 100875, China.
2. School of Finance, Shandong University of Finance and Economics, No.40, Shungeng Road, Shizhong District, Jinan, Shandong province 250014, China.

Yanran WU: bjfreelord@163.com
Xinlei ZHANG: Xinlei_Zh@163.com
Quanyi XU: xuquanyi@mail.bnu.edu.cn
Qianxin YANG: jadeyqx@163.com
Chao ZHANG: 20235290@sdufe.edu.cn

* Chao ZHANG is the corresponding author.

**Acknowledgements:** Yanran WU acknowledges that this work was supported by National Key R&D Program of China [grant numbers: 2023YFC3305401]

# Predicting Credit Spreads and Ratings with Machine Learning: The Role of Non-Financial Data

### Abstract

**We build a 167-indicator comprehensive credit risk indicator set**, integrating macro, corporate financial, bond-specific indicators, **and for the first time, 30 large-scale corporate non-financial indicators. We use seven machine learning models to construct a bond credit spread prediction model**, test their spread predictive power and economic mechanisms, and verify their credit rating prediction effectiveness. Results show **these models outperform Chinese credit rating agencies in explaining credit spreads**. Specially, **adding non-financial indicators more than doubles their out-of-sample performance vs. traditional feature-driven models.** Mechanism analysis finds **non-financial indicators far more important than traditional ones (macro-level, financial, bond features)—seven of the top 10 are non-financial** (e.g., corporate governance, property rights nature, information disclosure evaluation), the most stable predictors. Models identify high-risk traits (deteriorating operations, short-term debt, higher financing constraints) via these indicators for spread prediction and risk identification. Finally, **we pioneer a credit rating model using predicted spreads (predicted implied rating model)**, with full/sub-industry models achieving over 75% accuracy, recall, F1. This paper provides valuable guidance for bond default early warning, credit rating, and financial stability.

## 1. Introduction

Bond credit risk prediction and rating serve as the cornerstone for the accurate pricing and stable operation of the bond market. Traditionally, bond credit risk assessment has relied primarily on two types of prediction models: The first type is bond risk ratings derived from the internal rating models of rating agencies. Bonds with lower ratings usually indicate higher default risk. However, **a large body of literature argues that due to conflicts of interest arising from the "issuer-pays" business model and cognitive biases in analysts' prediction, credit ratings struggle to ensure objectivity and fairness. Issues such as lag, inflation, and predictive bias persist (Xia, 2014; Driskill et al., 2020; Li et al., 2021), making it difficult to effectively reveal bond credit risk in practice.**

The second type is quantitative models built on statistical data. Such models select risk prediction indicators based on prespecified functional forms, with typical examples including linear models such as the Altman Z-score model and ZETA model (Altman, 1968; Altman et al., 1977), as well as nonlinear models such as the Merton model (Merton, 1974) and KMV model. While these models ensure objectivity, both linear and non-linear variants require parameter estimation based on fixed variables. This leads to three key problems: the lack of dynamic time-variation in variable weight coefficients, insufficient diversity of predictive variables, and variable omission (Jiang et al., 2023). Moreover, as more variables with credit risk predictive power are incorporated into the model, the functional form becomes more complex. Linear models then face issues of "overfitting" and "the curse of dimensionality," while non-linear models suffer from unstable parameter estimation—both resulting in reduced out-of-sample predictive performance and efficiency.

In recent years, the financial sector has widely applied advanced technologies such as big data, artificial intelligence, and machine learning—and these technologies have demonstrated excellent performance in financial market prediction (Gu et al., 2020). Against this backdrop, **we attempt to apply machine learning methods to the bond market, propose and construct a set of credit risk prediction and rating models that are both objective, intelligent and dynamic** (Our overall research framework is shown in Figure 1). On one hand, machine learning models can make predictions based on real-time statistical data, which ensures the objectivity and timeliness of credit risk forecasts. On the other hand, machine learning models have distinct advantages in modeling complex high-dimensional data and variables: they possess strong data mining and pattern recognition capabilities, can automatically learn complex features and patterns from large volumes of data, and construct more accurate early warning models for bond default probability. These models also address issues such as "overfitting," "the curse of dimensionality," and unstable parameter estimation. Furthermore, when making predictions, they adopt rolling window or recursive window training methods to ensure that predictive variables obtain optimal weight coefficients in each period and resolve the lack of dynamic time-variation in variable weights and improve out-of-sample prediction accuracy.

**[Figure 1 inserts here]**

Before constructing our machine learning model, we first build a more forward-looking and timely credit risk prediction indicator set. Existing studies and mainstream rating agencies typically construct such credit risk prediction indicator sets from three dimensions: macroeconomic and policy levels, corporate financial levels, and bond feature levels (e.g., Bianchi et al., 2021; Bali et al., 2022; Bitetto et al., 2024; Jiang et al., 2023). However, traditional indicator sets have significant limitations. Macroeconomic and financial indicators mostly reflect historical static information, making it difficult to capture dynamic risk hazards in corporate operations (e.g., governance defects, information disclosure flaws). Bond-level indicators, on the other hand, focus on the attributes of the instruments themselves and overlook the underlying traits of firm as risk carriers.

With the advent of the big data era, massive volumes of data are emerging—unstructured corporate non-financial data such as media sentiment, ESG performance, corporate social responsibility reports, analyst reports, audit reports, and supply chain relationships are demonstrating stronger risk predictability (Bradshaw, 2011; Schneider, 2011; Fracassi et al., 2016; Gao et al., 2020; Li et al., 2021; Baker et al., 2022). This type of data can prematurely reveal potential risks like corporate governance failures and increased information asymmetry, compensating for the lag of traditional indicators. Yet existing studies have largely overlooked it. In view of this, we innovatively incorporate large-scale corporate non-financial characteristics into the indicator system, and build a credit risk prediction indicator set comprising 81 macro-level indicators, 47 corporate financial indicators (referencing the financial indicators used by rating agencies), 30 corporate non-financial indicators, and 9 bond features—totaling 167 indicators.

Next, considering that bond credit spreads combine objectivity, dynamism, and timeliness, and that there is a general pattern where a sharp rise in credit spreads usually accompanies the outbreak of credit defaults (both before and after), we therefore adopt bond credit spreads to measure bond credit risk. We select seven advanced machine learning models, including Random Forest (RF), AdaBoost Regression (AdaBosst), XGBoost Regression (XGBoost), Gradient Boosting Decision Tree (GBDT), LASSO Regression (LASSO), Ridge Regression (Ridge), and Elastic Net (Enet), and adopt both rolling window and recursive window training methods respectively to construct a machine learning-driven implicit bond credit spread prediction model (hereafter referred to as the "credit spread prediction model")[1].

Subsequently, we select all 11,317 corporate bonds, enterprise bonds, medium-term notes, and short-term financing bills issued by China's A-share listed firms between 2012 and 2023 as our sample to verify the effectiveness of the credit risk indicator system and the credit spread model. We choose China's bond market for three reasons: First, China's bond market has grown into the world's second-largest, yet scholars have not paid it attention matching its status. Instead, existing studies still focus mainly on developed markets in Europe and the United States, with greater emphasis on exploring bond pricing and return predictability (Bianchi et al., 2021; Bali et al., 2022; Bitetto et al., 2024). Second, as the world's largest emerging economy, China offers experience in building a credit spread prediction model based on its bond market—experience that provides significant reference value for developing similar models for other emerging economies. Third, global trade frictions and geopolitical factors have led to frequent credit default events in China's bond market in recent years. Wind data shows that in 2024, 15 new entities defaulted on or extended their bonds in China's domestic market, with the total scale of defaults and extensions reaching RMB 59.431 billion. This means institutional investors now face a critical issue in global asset allocation and risk management: accurately predicting and timely preventing credit risk in China's bond market.

---

[1] The credit spread prediction model constructed in this study uses current-month data of predictors to predict the bond credit spreads for the next month. In other words, we predict the implicit spreads of bonds for the next month one month in advance.

To evaluate the model's predictive performance, we use traditional credit rating-based spread prediction models as the comparison benchmark, as some Chinese studies have shown that credit rating-based spread prediction models outperform statistical models like the Altman Z-score in predictive performance (Jiang et al., 2023). We then follow the method of Gu et al. (2020) and calculate the out-of-sample predictive goodness of fit ($R^2_{OS}$) for evaluating the incremental performance of machine learning models compared to traditional credit rating-based spread prediction models. Our empirical results show that compared to traditional credit rating-based spread prediction models, the machine learning-driven credit spread prediction model predicts bond credit spreads more accurately and timely. It specifically exhibits lower mean absolute error (MAE) and root mean squared error (RMSE), as well as higher out-of-sample goodness of fit ($R^2$) and out-of-sample predictive goodness of fit ($R^2_{OS}$). Notably, after incorporating corporate non-financial indicators into machine learning models, we further find that the predictive performance improves by more than double on average compared to machine learning models driven by traditional feature indicators. Among these models, the RF model and XGBoost model exhibit the best credit spread prediction performance.

We could also intuitively demonstrate our machine learning model prediction performance through randomly selected prediction case. Figure 2 shows the credit spread trends of two randomly selected Vanke[2] bonds—22 Vanke MTN005 and 22 Vanke GN001 (marked by the red line in Figure 2 below)—and the predicted credit spread trends of these two bonds generated by our real estate industry credit spread model constructed using the RF model (marked by the blue line in Figure 2 below). By early October 2023, the market credit spreads of 22 Vanke MTN005 and 22 Vanke GN001 have already signaled a sharp rise in credit risk (red circle in Figure 2). However, international rating agencies such as Fitch and Moody's did not downgrade the bonds' ratings (from Baa1 to Baa3) until October 17 and November 24, 2023, respectively.

---

[2] Vanke Group is a leading urban-rural construction and lifestyle service provider in China, with businesses spanning residential development, property services, long-term rental apartments, logistics and warehousing, and has been ranked the world's most valuable real estate brand for consecutive years.

S&P maintained a BBB+ rating for the bonds throughout 2023 and made no downgrades. Meanwhile, China's nine major rating agencies, including CCXI and Lianhe Credit Rating, also kept an AAA rating for the bonds the entire year. Neither domestic nor international rating agencies downgraded their ratings in a timely manner, and the performance of these rating agencies fully reveals the lag in the current rating system. By contrast, we can intuitively find the excellent performance of our machine learning prediction model from the blue line in Figure 2. By the end of September 2023 (blue circle in Figure 2), our real estate industry credit spread model—constructed using the RF model—had already predicted that the credit spreads of the two Vanke bonds would rise sharply in October and that credit default risk would increase significantly. Our model warns credit risk far ahead of a host of rating agencies, including Fitch, Moody's, Standard & Poor's, and nine major rating agencies in China.

**[Figure 2 inserts here]**

One potential problem with machine learning models is the difficulty of identifying the internal decision-making logic behind their predictions. To address this, we use the SHAP  (SHapley Additive exPlanations) method to identify the importance of indicators in credit spread prediction, so as to determine which indicators matter during the prediction. As an advanced model interpretability technique, SHAP can penetrate the "black box" of complex models: it accurately attributes prediction results to each input feature indicator and reveals the direction and intensity of how specific features drive credit spread prediction values. Furthermore, logically, when corporate bond credit risk rises, it typically involves processes such as deteriorating corporate operations, increasing leverage, short-termized debt structures, and tighter corporate financing constraints. In more severe cases, corporate financing pressure surges sharply, triggering the outbreak of debt default risk. Based on this, we reference Jiang et al. (2023) and use portfolio analysis to further reveal the internal decision-making mechanism of machine learning models—by examining how well these models can predict this aforementioned process.

Our empirical results show that compared to traditional indicators (at the macro, corporate financial, and bond levels), non-financial indicators are significantly more important: 7 out of the top 10 indicators by importance are non-financial. In particular, corporate non-financial indicators—such as corporate governance status, nature of property right, and information disclosure evaluation—serve as the most important and stable predictive signals for machine learning models. Additionally, further portfolio analysis verifies the decision-making mechanism mentioned before: by comprehensively utilizing these indicators, machine learning models effectively identify high-risk features (e.g., deteriorating corporate operations, short-termized debt structures, and tighter corporate financing constraints), thereby achieving the identification and prediction of credit risk.

Finally, since credit spreads implicitly contain bond credit ratings, we could also use credit spread prediction models to forecast implicit bond credit ratings. To do this, we develop a credit rating model—leveraging implicit ratings from market credit spreads, the credit spread prediction model, and the RF and XGBoost models. We name our model the "implicit credit rating prediction model based on credit spreads" (hereafter "implicit rating prediction model") and demonstrated its strong performance in credit rating prediction. Our implicit rating prediction model delivers over 80% accuracy across the entire industry and over 70% for sub-industries. Notably, it achieves the highest accuracy (over 83%) for the financial and real estate sectors. Additionally, we use the SHAP method for further mechanism analysis and find that corporate non-financial indicators—such as corporate governance status, nature of property right, information disclosure evaluation, and registered capital—also play a key role.

Importantly, our results strongly show that non-financial features (e.g., corporate governance, nature of property right) are critical for credit spread prediction and credit rating, yet traditional market rating agencies apparently underestimate their impact. For instance, though S&P China also consider corporate governance in its rating process, it fails to integrate it effectively with rating indicators. Instead, it typically only discloses corporate governance status or issues risk warnings based on financial indicator ratings (e.g., adjusting ratings by notches, with no more than one level of impact)—limiting the effectiveness of its spread prediction and ratings. Using machine learning, we not only integrate this critical information effectively and propose a machine learning-driven credit rating prediction model with non-financial data, but also significantly boost the effectiveness of credit spread and credit rating prediction.

Our marginal contributions are reflected in at least the following two aspects: **on**

**the one hand, we propose a more forward-looking credit risk prediction indicator set that aligns with the characteristics of emerging economies. In particular, we incorporate a large-scale set of non-financial information (30 indicators) for the first time, and significantly improve the predictive performance of the machine learning-driven credit spread model.** Compared with existing studies that use machine learning to predict bond return in the developed economies (Bianchi et al., 2021; Bali et al., 2022; Zhai et al., 2024), we develop a suitable indicator set for predicting bond credit risk in emerging economies, enabling the prediction of bond credit spreads and implicit ratings in the Chinese market. More importantly, our indicator set not only covers traditional credit spread prediction indicators based on corporate financial indicators, but also innovatively achieves large-scale quantification of corporate non-financial data and incorporates a broader range of macro-policy data (e.g., monetary policy, monetary policy uncertainty), thus constructing a more comprehensive and forward-looking credit spread prediction indicator set.

Specifically, we highlight the importance of non-financial indicators (such as corporate governance, nature of property right, and information disclosure evaluation) in credit risk assessment. The inclusion of these non-financial indicators enables the model's predictive performance to not only outperform the ratings of traditional credit rating companies, but also—compared with machine learning models that only include characteristic indicators at the macro, corporate financial, and bond feature levels—its out-of-sample predictive performance further improved by more than double on average. This means that through the systematic optimization of the credit spread prediction indicator set—especially the integration of large-scale corporate non-financial indicators into the system **(7 out of the top 10 credit indicators in terms of model importance are non-financial indicators)—we can significantly outperform and improve the existing credit spread prediction models of credit rating agencies,** ultimately matching the predictive performance of cutting-edge machine learning models trained on traditional indicators.

**On the other hand, for the first time, we develop a more objective and forward-looking credit rating method for China's securities market.** We develop a credit spread prediction technology with the characteristics of objectivity, forward-looking, and localization, which enables us to achieve dynamic and accurate identification of credit spreads. We also uncovered the economic mechanisms behind this model. Building on this foundation, **we develop a credit rating model based on credit spread classification and machine learning methods—the "implicit credit rating prediction model based on credit spreads"—and demonstrated its outstanding performance in credit rating.** This model can overcome the conflicts of interest of rating agencies and the cognitive biases of credit analysts (Xia, 2014; Driskill et al., 2020; Li et al., 2021), providing a more transparent and efficient credit risk classification and rating method for China's securities market. This helps enhance global investors' ability to identify and prevent credit risks in China's bond market, and reduces losses caused by credit defaults.

We organize the remainder sections as follows: Section 2 conducts a literature review. Section 3 introduces the research design and data selection. Section 4 presents the empirical results of the credit spread prediction model. Section 5 discusses and analyzes the economic mechanisms behind the model. Section 6 examines the economic value of the credit spread prediction model in credit rating. We conclude in Section 7.

## 2. Literature review

### 2.1 Indicators for bond credit risk prediction

Existing studies has conducted extensive explorations regarding the predictive indicators of bond credit risk (or the influencing factors of credit risk) based on three perspectives: macro-level feature, firm-level feature, and bond-level feature.

#### 2.1.1 Macro-level indicators

At the macro level, early studies focus more on the predictive effects of macroeconomic factors—such as GDP growth rate, industrial value-added, CPI, interest rates, economic cycles, stock index returns, financial market volatility, and regional economic conditions—on bond credit risk (Merton, 1974; Altman, 1998; Hull & White, 1994; Collin-Dufresne et al., 2001; Lucas et al., 2003; Duffie et al., 2007; Wu & Zhang, 2008; Tan & Yan, 2010). Subsequent studies, however, have found that macro policies are also key factors for early warning of bond credit risk, as they transmit between money markets and bond markets, thereby exerting a significant impact on the credit risk of corporate bonds. For instance, Zhu (2013) finds that the volatility and uncertainty of monetary policy affect market liquidity risk and market expectations, which in turn influence the credit spreads of bonds. Furthermore, in some recent studies, scholars have begun to focus on the pricing effects of government public data (such as transport volume data, environmental data) in bond markets and stock markets (Du et al., 2024).

### 2.1.2 Firm-level indicators

At the firm level, early studies primarily focus on the predictabilities of corporate financial features for bond credit risk. For example, the traditional Altman Z-score model constructs a credit risk prediction system by selecting 5 indicators covering aspects such as short-term solvency, profitability, cumulative profitability, market value-to-debt coverage ratio, and operational efficiency. Later, the ZETA model further added 2 additional indicators—including the interest coverage ratio (to measure long-term solvency) and earnings stability. Molina (2005) argued that cash flow levels exert a significant impact on credit default risk. Drawing on research on stock market anomalies, Jiang et al. (2023, 2024) constructed 70 firm characteristic variables across six categories: valuation and growth, investment-related, profitability-related, momentum-related, trading friction-related, and intangible asset-related. Most of these variables are financial indicators, and the study confirmed that metrics such as the current ratio, market-to-book ratio, change rate of fixed assets and inventory, firm return rate, and EBIT (earnings before interest and taxes) play a crucial role in identifying high-risk bonds using machine learning and deep learning models (Jiang et al., 2024).

In recent years, with the emergence of massive volumes of data, the academic and industrial communities have paid greater attention to the incremental effect of corporate non-financial information (beyond firms' financial data) on credit risk prediction—and have found that certain firm-related non-financial information can be used to improve the accuracy of credit risk prediction. Some studies have shown that corporate governance factors show predictive power for firms' credit risk: for instance, ownership structure (Chiang et al., 2015), board characteristics (Ashbaugh-Skaife et al., 2006), executive characteristics (Bonsall et al., 2017), and information disclosure quality (Fang et al., 2013) all significantly influence default risk. Additionally, some scholars have found that firms' operational characteristics—such as nature of property of ownership (Gao & Zhao, 2021), corporate innovation (Griffin et al., 2018), supply chain conditions (Zhang et al., 2015), corporate social responsibility (CSR) performance (Oikonomou et al., 2014; Jiraporn et al., 2014), environmental performance (Bauer & Hann, 2010), and ESG ratings (Schneider, 2011)—are helpful for predicting firms' credit risk. Particularly with the development of text analysis technology, scholars have discovered that text-based information (e.g., analyst reports (Bradshaw, 2011), annual reports of listed companies (Jiang et al., 2019), audit reports (Chen & Li, 2013), and media sentiment (Gao et al., 2020; Baker et al., 2022) contains rich insights for credit risk assessment.

### 2.1.3 Bond-level indicators

At the bond-level feature, existing studies find that bond-related features such as restrictive covenants (Bradley & Roberts, 2015), remaining maturity (Wang et al., 2012), and cross-default clauses (Yang & Jiang, 2012) are directly associated with default probability and credit risk. Studies by Jiang et al. (2023, 2024) have further confirmed that feature indicators like bond credit ratings also play an important role in credit risk models driven by machine learning and deep learning.

In summary, existing research on the predictive factors of bond credit risk is relatively comprehensive, covering multiple dimensions including the macro level, firm financial and non-financial levels, and bond level. In particular, corporate non-financial features are demonstrating strong risk prediction performance. This has laid a research foundation for us to construct a credit risk prediction indicator system that is more in line with emerging market (like China) characteristics and more forward-looking.

## 2.2 Machine learning and bond credit risk assessment model

Traditionally, bond credit risk assessment primarily relies on two types of prediction models. One is the bond risk rating derived from the internal rating models of rating agencies: bonds with lower credit ratings typically carry higher credit default risk. However, under the "issuer-pays" business model, the objectivity and effectiveness of credit ratings issued by rating agencies have been questioned, with problems such as inflated credit ratings and delayed adjustments (Xia, 2014). At the same time, when rating analysts conduct risk grading, they need to make subjective inferences and predictions based on data. During this process, analysts usually focus on limited specific data, which easily leads to cognitive biases such as limited attention and anchoring effect, resulting in rating distortions (Driskill et al., 2020; Li et al., 2021). Notably, during periods of economic prosperity, analysts generally tend to exhibit "rating inflation," while during economic downturns, they often "overreact." This procyclical rating decision-making makes it difficult for rating results to objectively reflect true credit conditions of firms. Especially in recent years, frequent credit default events of high-rated bonds have sparked widespread controversy over the effectiveness of bond credit ratings, and the reliability of bond ratings as an indicator of credit risk has been seriously questioned.

The other type of credit risk prediction models is constructed by selecting credit risk predictive indicators, based on statistical data and specific functional forms. Classified by their functional forms, these models can be divided into linear and non-linear credit risk prediction models. Linear models primarily predict credit risk by performing linear regression, Probit regression, or Logit regression between credit risk and several indicators to obtain estimated coefficients. Examples include the traditional Altman Z-score model and ZETA model (Altman, 1968; Altman et al., 1977). Representative non-linear models, on the other hand, mainly include the Merton model (Merton, 1974) and the KMV model.

However, regardless of whether they are linear or non-linear, these models all face challenges related to parameter estimation. This leads to two key issues: first, the lack of dynamic time-variation in variable weight coefficients, and second, insufficient diversity in predictive variables—along with the potential for variable omission (Jiang et al., 2023). Meanwhile, as more variables with credit risk predictive power are identified, the introduction of high-dimensional variables further complicates the model's functional form. Linear models may then encounter problems of "overfitting" and the "curse of dimensionality," while traditional non-linear models may suffer from unstable parameter estimation. Both scenarios result in a decline in out-of-sample predictive performance and efficiency.

In recent years, the widespread application of technologies such as machine learning in the financial prediction (Gu et al., 2020; Zhao & Cheng, 2022) has provided methodological support for the assessment of bond credit risk. Machine learning models can make predictions based on real-time statistical data, ensuring the objectivity and timeliness of credit risk predictions. Meanwhile, when dealing with complex high-dimensional data and variable modeling, machine learning methods possess stronger data mining and pattern recognition capabilities: they automatically learn complex features and patterns from large volumes of data, construct more accurate early warning models for bond default probability, and address issues such as "overfitting," "the curse of dimensionality," unstable parameter estimation, and the lack of dynamic time-variation in variable weight coefficients—ultimately helping to improve out-of-sample prediction accuracy.

Therefore, some scholars have begun to apply machine learning methods to research on credit risk in the bond market. For instance, early studies by Bianchi et al. (2021) and Bali et al. (2022) applied machine learning methods to the prediction of U.S.

Treasury bond yields and corporate bond yields, respectively, and validated their effectiveness. Subsequent literature further verified the predictive power of machine learning models for bond risk premiums (Zhai et al., 2024), as well as their predictive power for credit risk in multiple specific fields, including small and micro-enterprises, bank credit, auto loans, online credit, and supply chain finance (Bitetto et al., 2023; Liu et al., 2022; Chang et al., 2024; Gasmi et al., 2025; Lin et al., 2025). Furthermore, a small number of Chinese scholars have validated the feasibility and effectiveness of machine learning methods for bond credit risk prediction based on China's bond market (Jiang et al., 2023) and test the predictive power of more advanced deep learning models (such as CDL and FAAN) for credit risk (Jiang et al., 2024; Wang et al., 2025). However, their machine learning models are mainly based on traditional macro-level, corporate financial-level, and bond-level indicators, and lack the introduction of indicators that are more in line with the features of emerging market economies and more forward-looking.

In summary, although a small number of existing studies have preliminarily verified the effectiveness of machine learning in credit risk prediction, they still lack broader micro-level evidence support based on the bond markets of emerging market economies such as China. Additionally, there is a scarcity of research exploring the effectiveness of machine learning in credit rating. Furthermore, in studies conducted in China, there is an overemphasis on the application of machine learning methods (e.g., Jiang et al., 2024), with insufficient attention to the systematic construction of credit risk prediction indicators. Chinese scholars primarily construct models by selecting traditional credit risk prediction indicators from the macro, corporate financial, and bond levels—drawing on research or experience from the bond or stock markets of developed countries—while lacking the introduction of more localized and forward-looking indicators. With the advent of the big data era, massive volumes of data are emerging. The integration of this data—especially the quantification and construction of a large number of corporate non-financial indicators—is expected to drive a significant leap in the predictive performance of existing machine learning models. Finally, existing machine learning models provide insufficient explanation of their underlying economic mechanisms. Most studies remain limited to comparing predictive performance and lack in-depth analysis of the economic mechanisms behind machine learning's ability to identify credit risk.

In view of the limitations of the existing studies, we attempt to incorporate a large amount of corporate non-financial feature data into the indicator database, construct a credit risk indicator system that is more in line with the characteristics of emerging market economies and more forward-looking, further test and enhance the performance of machine learning-driven credit risk prediction models, systematically explore the economic mechanisms underlying machine learning models, and attempt to develop a machine learning-based credit rating prediction model.

## 3. Data and Methodology

### 3.1 Sample and data interpretation

We collect a total of 11,317 corporate bonds, enterprise bonds, medium-term notes, and short-term financing bills—all issued by China's A-share listed companies between 2012 and 2023—as our research samples. We source bond credit rating data from the Wind database, mainly obtain corporate financial data from the China Stock Market and Accounting Research (CSMAR) database, draw bond and macro-level data from the Choice database[3], and collect corporate non-financial data from the CSMAR database, Chinese Research Data Services Platform (CNRDS) database, and Sino-Securities ESG Rating database[4].

Referring to existing studies, we process the data as follows: (1) We uniformly convert all data to a monthly frequency; for financial data, we fill in missing values with the monthly median of the corresponding indicator. (2) We determine the lead-lag relationships of variables based on the actual release time of each dataset, then sequentially match bond rating data with bond feature data, financial and non-financial features data of issuing firms, and macro-level data, and exclude samples with missing bond ratings, missing bond features, or missing corporate non-financial features data.

---

[3] Choice database is an intelligent financial data brand under Oriental Fortune and a leading financial data service provider in China.

[4] The Sino-Securities ESG Rating database is compiled by Sino-Securities Index Information Service (Shanghai) Co., Ltd., a leading ESG rating agency in China that specializes in comprehensive services pertaining to indices and index-based investments

(3) We adjust macro-level indicators, corporate financial indicators, and corporate non-financial indicators to lag behind credit spread indicator by 1 month[5], enabling us to make 1-month-ahead predictions. (4) We winsorize all continuous variables at the 1% level and uniformly standardize the data using Z-score (resulting in a mean of 0 and a standard deviation of 1).

After the above processing, we obtain a final sample containing 2,636 bonds and a total of 45,283 monthly observations. It should be noted that machine learning models require training periods: when we take January 2012 as the starting point and use 6-month as training window, the corresponding intervals for benchmark empirical results are July 2012–December 2023.

## 3.2 Machine learning-driven credit spread prediction model

We take bond credit spread as the proxy variable for credit risk, and construct the credit spread prediction model as follows:

$$Credit\ spread_{i,t} = f(M_{i,t-1}, F_{i,t-1}, NF_{i,t-1}, B_{i,t-1}) + u_{i,t} \qquad （1）$$

where the subscript $i$ denotes a bond, and $t$ denotes a time period. $Credit\ spread_{i,t}$ indicates the credit spread of bond $i$ in period $t$. Following the approach of Jiang et al.

---

[5] To simulate a real-world prediction environment, for data at all levels in a given month, we use the latest released data available before 6:00 PM on the last day of that month. For instance, when predicting the credit spread at the end of November 2022 based on data at the end of October 2022, we use the latest data accessible before 6:00 PM on October 31, 2022. Specifically, we adopt Q3 2022 financial report data (all of which had been released by October 31) for financial indicators, while monthly macroeconomic data refers to the September data released in October. For data with a frequency of quarterly (3-monthly) or annual (12-monthly), we use the latest released data as of the end of July 2022 and the end of 2021 respectively.

(2023), we match the ChinaBond Treasury bond yield curve data with the same remaining maturity to bond $i$, and calculate the monthly credit spread of bond $i$ by computing the difference between the monthly yield to maturity of bond $i$ and the yield to maturity of the corresponding ChinaBond Treasury bond curve. $M_{i,t-1}$、$F_{i,t-1}$、$NF_{i,t-1}$、$B_{i,t-1}$ denote macro-level variables, corporate financial variables, corporate non-financial variables, and bond-level variables, respectively. $f(\cdot)$ denotes the selected machine learning algorithm. Referring to the methods of Gu et al. (2020) and Jiang et al. (2023), we select seven machine learning models—including Random Forest (RF), XGBoost Regression (XGBoost), AdaBoost Regression (AdaBoost), Gradient Boosting Decision Tree (GBDT), LASSO Regression (LASSO), Ridge Regression (Ridge), and Elastic Net (Enet)—to construct the bond credit spread prediction model.

Random Forest (RF) model selects samples via the Bootstrap method and randomly chooses a subset of variables during training to build the model. This approach effectively reduces the correlation between decision trees, decreases prediction variance, and thereby improves model accuracy. During model training: first, B random training sets are constructed using Bootstrap sampling, and the corresponding B decision trees are trained by minimizing the Gini coefficient. Next, when building split nodes for the trees, q variables are randomly sampled without replacement from m influencing factors (with the default value of q being m/3), and split nodes are identified from these q influencing factors. Finally, the B random training sets are used for training, and the node with the most votes is selected as the final node.

The XGBoost Regression (XGBoost) model is based on a gradient boosting framework and regularization optimization, making it suitable for high-dimensional sparse feature scenarios in credit risk prediction. During training, guided by the negative gradient of the loss function, regression trees are built one by one to fit residuals. Meanwhile, L1/L2 regularization is introduced to constrain tree complexity and avoid overfitting. For credit rating data, its advantages include: supporting automatic missing value handling (skipping missing feature branches via a sparse-aware algorithm), which adapts to common missing value scenarios in financial indicators; using second-order Taylor expansion to accelerate gradient descent, resulting in significantly higher training efficiency than traditional gradient methods when handling interaction effects between macroeconomic and corporate features; and an regularization mechanism that effectively reduces model fluctuations caused by industry heterogeneity.

In addition, AdaBoost Regression (AdaBoost), Gradient Boosting Decision Tree (GBDT), LASSO model (LASSO), Ridge Regression (Ridge), and Elastic Net (Enet) are also mainstream machine learning models in existing research. See Appendix A for specific details on their model characteristics.

### 3.3 Indicator set for credit spread prediction model

We, for the first time, try to incorporate a large-scale set of 30 non-financial indicators into the credit prediction indicator set, constructing a total of 167 credit risk prediction indicators covering four dimensions: macro-level, corporate financial level, corporate non-financial level, and bond feature level.

For macro-level predictive indicators ($M_{i,t-1}$), we select 81 indicators across 10 dimensions, including regional environmental indicators, macroeconomic factors, resources, agriculture, environment, fixed investment, passenger volume, passenger turnover, freight volume, and freight turnover. Notably, **existing literature focuses on macroeconomic performance of the real economy such as GDP and fiscal conditions—for example, the 15 macro indicators in Jiang et al. (2023)—while overlooking the impact of monetary policy on credit risk.** Considering that monetary policy and the resulting market liquidity form the basis of interest rate pricing, we include variables related to monetary policy (e.g., money supply) in the macroeconomic factor variables.

For corporate financial indicators ($F_{i,t-1}$), we select 47 indicators across 11 dimensions—profitability, leverage, firm size, financial coverage, liquidity, asset structure, growth, R&D and innovation, cash flow, operational capacity, and valuation—primarily based on existing literature (Gu et al., 2020; Jiang et al., 2023, etc.) and the indicator set of mainstream rating agencies.

Regarding the corporate non-financial indicators ($NF_{i,t-1}$), non-financial information—such as internal firm conditions (Oikonomou et al., 2014; Jiraporn et al., 2014), external public opinion environment (Gao et al., 2020; Baker et al., 2022), supply chain (Zhang, 2015), external support (Chen & Li, 2013), and corporate governance performance (Ashbaugh-Skaife, 2006; Chiang et al., 2015)—has demonstrated significant predictive power in the bond market. However, existing literature on credit risk prediction overlooks the potential benefits of these indicators in improving machine learning performance. Thus, we select 30 indicators across 5 dimensions: internal firm conditions, external firm environment, supply chain, external firm support, and corporate governance. The "internal firm conditions" dimension primarily reflects a firm's basic situation and operational status (e.g., registered capital, information disclosure evaluation), measuring the transparency, standardization, and sustainability of its internal management and operations. The "external firm environment" dimension reflects a firm's performance in social public opinion and market environment (e.g., news exposure, proportion of positive/neutral/negative media coverage). The "supply chain" dimension reflects the stability of a firm's upstream and downstream relationships (e.g., customer concentration). The "external support" dimension reflects the resource support a firm obtains from outside (e.g., government subsidies, political connections). The "corporate governance" dimension reflects a firm's governance status (e.g., shareholding structure of shareholders and the board of directors).

For bond characteristic indicators ($B_{i,t-1}$), we select 9 predictive indicators based on the studies of Bradley & Roberts (2015), Wang et al. (2012), and Jiang et al. (2023), including bond issuance maturity, remaining maturity, trading venue, issuance scale, and guarantee status.

See Table 1 and Appendix B for the credit risk prediction indicators and their calculation methods across all dimensions. Descriptive statistics of all variables are provided in Appendix C.

**3.4 Model training and evaluation methods**

**3.4.1 Model training method**

Following existing studies, we adopt two training methods—"rolling window" (Jiang et al., 2023) and "recursive window" (Gu et al., 2020)—to divide the training set and validation set, and conduct parameter selection. The former better captures the time-varying nature of parameters, while the latter has the advantage of accounting for the long-memory property of sequences. Specifically, for the rolling window prediction method, we set M as the length of the training interval. When predicting the bond credit spread at a given time, we use the M months preceding the current period (i.e., a total of M months from t-M to t-1) as the training set and validation set. We then determine the optimal parameter values for the current period and the monthly credit spread predictions for all bonds, following the criterion of minimizing the Root Mean Squared Error (RMSE) of the validation set. In other words, the training set of "rolling window" rolls forward gradually with the training interval, thus maintaining a fixed length. In the benchmark analysis, we set M = 6, and use M = 3 and M = 12 for robustness tests.

For the recursive window method, the length of the training set gradually increases. Specifically, assuming M = 6, the initial training interval is January 2012–June 2012, where the initial training set is January 2012–April 2012 and the initial validation set is May 2012–June 2012. After the first round of training, the training interval increases by one month to seven months: the training set becomes January 2012–May 2012 (i.e., its length gradually increases), while the validation set maintains a fixed length and rolls forward by one month to June 2012–July 2012. Similarly, we obtain the optimal parameter values for the current period and the monthly credit spread predictions for all bonds by minimizing the RMSE of the validation set. In this case, the length of the training set for "recursive window " gradually increases. In the benchmark analysis, we set M = 6, and verify it with M = 3 and M = 12 in robustness tests.

### 3.4.2 Model evaluation method

As in Gu et al. (2020), we construct the out-of-sample predictive $R^2$ ($R_{OS}^2$) to compare the advantages of our machine learning-based credit spread prediction model over traditional credit risk early warning models and analyze its improvement in credit spread predictive performance. $R_{OS}^2$ is used to assess the "incremental" predictive power of our indicator set relative to traditional bond credit risk early warning models and to measure the effectiveness of the indicators. The formula for the out-of-sample predictive $R_{OS}^2$ is as follows:

$$R_{OS}^2 = 1 - \frac{\sum_{t=M+1}^{T} \sum_{i=1}^{N_t} \left(\text{Credit spread}_{i,t} - \widehat{\text{Credit spread}}_{i,t}\right)^2}{\sum_{t=M+1}^{T} \sum_{i=1}^{N_t} \left(\text{Credit spread}_{i,t} - \widetilde{\text{Credit spread}}_{i,t}\right)^2} \qquad （2）$$

where $\widehat{\text{Credit spread}}_{i,t}$ denotes the credit spread predicted by the machine learning model, and $\widetilde{\text{Credit spread}}_{i,t}$ denotes the credit spread predicted by the traditional credit risk prediction model. Based on the findings of Jiang et al. (2023), among traditional credit spread prediction models, credit rating models exhibit stronger out-of-sample predictive power than other traditional models such as the Altman Z-score and Altman's five-factor model. Therefore, we use a univariate linear regression constructed based on aggregated ratings from nine major credit rating agencies in China as the benchmark model[6] to evaluate the predictive power of our machine learning-based credit spread model. If $R_{OS}^2 \leqslant 0$, it indicates that the machine learning models are ineffective. If $0 < R_{OS}^2 < 1$, it means the predictive performance of our machine learning model is superior to that of the credit rating model, that is the models are effective. The higher the $R_{OS}^2$, the stronger the predictive power of the models and selected indicators.

## 4. Empirical Results and Analysis

### 4.1 Machine learning model prediction error and out-of-sample predictability

In this section, we focus primarily on the prediction errors and out-of-sample predictive $R^2(R_{OS}^2)$ of various machine learning-driven credit spread prediction models. For each machine learning model, we conduct training using rolling and recursive windows. The initial training set starts in January 2012, with a 6-month training period, and the corresponding interval for benchmark empirical results is July 2012 to December 2023. Descriptive statistics for the 81 macroeconomic variables, 47 corporate financial variables, 30 corporate non-financial variables, and 9 bond-specific variables included in the machine learning models are presented in Appendix C.

---

[6] The nine major credit rating agencies in China include: (1) Dagong Global Credit Rating Co., Ltd.; (2) Oriental Jincheng International Credit Rating Co., Ltd.; (3) Lianhe Credit Rating Co., Ltd. (4) Lianhe Credit Information Service Co., Ltd.; (5) Pengyuan Credit Rating Co., Ltd.; (6) Shanghai Brilliance Credit Rating & Investors Service Co., Ltd.; (7) Shanghai Far East Credit Rating Co., Ltd. (8) CCXI Credit Rating Co., Ltd.; (9) CCXI Securities Rating Co., Ltd.

### 4.1.1 Prediction results of credit spread model with rolling window training

Table 2 reports the Mean Absolute Error (MAE), Root Mean Squared Error (RMSE), out-of-sample $R^2$ ($R^2$), and out-of-sample predictive $R^2$ ($R^2_{OS}$) of various machine learning models under the rolling window training method—with one group using traditional feature indicators (including only macro-level, corporate financial-level, and bond-level indicators) and the other incorporating corporate non-financial feature indicators.

**[Table 2 inserts here]**

From the results of models based on traditional feature indicators, the out-of-sample predictive power of machine learning-based credit spread prediction models is generally superior to that of the benchmark model. This is reflected in the following: the RMSE of all machine learning models is lower than that of the benchmark model; the MAE of 6 out of 7 models is lower than that of the benchmark model; and the $R^2$ of 6 models is higher than that of the benchmark model. Furthermore, as a core evaluation criterion, the average $R^2_{OS}$ of the seven machine learning models is 0.117, and 6 out of the 7 models have a positive $R^2_{OS}$. **In particular, RF and XGBoost models achieve the best out-of-sample predictive $R^2$, with $R^2_{OS}$ values of 0.296 and 0.247 respectively. This indicates that even when using the same traditional feature data, machine learning models still outperform the benchmark model.** Since the comparison benchmark in our study is the credit risk model commonly used by domestic rating agencies in China, this implies that the credit spread model we constructed has surpassed the predictive power of market rating agencies for credit spreads.

More importantly, **after incorporating corporate non-financial feature indicators into the traditional feature indicators, the predictive power of the machine learning models improves significantly:** the MAE and RMSE of all machine learning models decrease, while their $R^2$ and $R_{OS}^2$ increase. Additionally, all models have a positive $R_{OS}^2$, and their MAE, RMSE, and $R^2$ are all better than those of the benchmark model. Meanwhile, **after the introduction of non-financial indicators, the RF and XGBoost models still maintain the best out-of-sample predictive $R^2$, with their $R_{OS}^2$ values rising sharply from the original 0.296 and 0.247 to 0.539 and 0.547 respectively—representing increases of 82.09% and 121.46%.** In terms of average results, after incorporating non-financial indicators, the average MAE and RMSE of all models decrease from 1.007 and 1.465 to 0.861 and 1.266 (a reduction of 14.50% and 13.58% respectively), while the average $R^2$ and $R_{OS}^2$ increase from 0.266 and 0.117 to 0.496 and 0.393 (a growth of 86.47% and 235.90% respectively).

These results not only demonstrate that machine learning-based credit spread prediction models outperform traditional credit risk models, but also indicate that machine learning models exhibit stronger predictive power when corporate non-financial indicators are integrated into traditional feature indicators. This fully verifies the effectiveness of incorporating corporate non-financial indicators into traditional feature indicators to construct our credit risk prediction indicator set and credit spread prediction model—with the inclusion of non-financial indicators significantly enhancing the predictive performance of machine learning models for credit spread. At the same time, since the comparison benchmark is the credit risk model commonly used by China's rating agencies, the credit spread model we constructed based on machine learning methods and traditional indicators is already more effective than existing rating agencies' models. Needless to say, its predictive performance is further improved after incorporating non-financial indicators, which provides a more effective rating method for Chinese bond market credit rating.

### 4.1.2 Prediction results of credit spread model with recursive window training

Corresponding to Table 2, Table 3 reports the MAE, RMSE, $R^2$, and $R_{OS}^2$ of various machine learning models under the recursive window training method—with one group using traditional feature indicators (including only macro-level, corporate financial-level, and bond-level indicators) and the other incorporating corporate non-financial feature indicators.

Similar to the results in Table 2, the out-of-sample predictive power of machine learning-driven credit spread prediction models is significantly superior to that of the benchmark model when using traditional feature indicators. This is evidenced by the following: the RMSE of all models is lower than that of the benchmark model; the MAE of 6 out of 7 models is lower than that of the benchmark model; and the $R^2$ of 6 models is higher than that of the benchmark model. In terms of the core evaluation criterion, the average $R_{OS}^2$ of all models is 0.201, and 6 out of the 7 models have a positive $R_{OS}^2$. The RF and XGBoost models again achieve the best $R_{OS}^2$, which further validates the results in Table 2.

At the same time, the results in Table 3 show that after incorporating corporate non-financial feature indicators into the traditional feature indicators, the predictive power of the machine learning models improves significantly. This is reflected in the decrease in MAE and RMSE, and the increase in $R^2$ and $R_{OS}^2$ for all machine learning models. **Among these, the RF and XGBoost models— which perform best in out-of-sample prediction—see their $R_{OS}^2$ values rise from 0.438 and 0.449 to 0.620 and 0.638 respectively, representing increases of 41.55% and 42.09%.**

Finally, in terms of average results across all models: the average MAE and RMSE decrease from 0.974 and 1.443 to 0.852 and 1.273 (a reduction of 12.53% and 11.78% respectively), while the average $R^2$ and $R_{OS}^2$ increase from 0.336 and 0.201 to 0.513 and 0.413 (a growth of 52.68% and 105.47% respectively). The results in Table 3 further verify the superiority of machine learning-based credit spread prediction models over traditional credit risk model, as well as the crucial role of corporate non-financial indicators in enhancing the credit spread predictive performance of machine learning models.

### 4.1.3 Performance comparison of various credit spread prediction models

Furthermore, we use the modified Diebold-Mariano (DM) statistic supported by Gu et al. (2020) to test the differences in out-of-sample credit spread predictive performance among various machine learning models. Specifically, based on the machine learning results trained with a rolling 6-month window, we calculate the modified DM statistic for each pair of models, and the results are presented in Table 4. As the results shown in Table 4, **the RF model and the XGBoost model are the best-performing models among the 7 machine learning models.** At the 5% significance level and after applying the conservative Bonferroni adjustment, the credit spread predictive power of these two models is significantly superior to that of the other machine learning models, **while there is no statistically significant difference in predictive performance between the two models themselves.**

**[Table 4 inserts here]**

## 4.2 Robustness tests

To check the robustness of our machine learning-driven spread prediction models, we conducted a series of robustness tests. First, we check the model's predictive performance by using different training window periods (M = 3, 12). Second, we further check the predictive performance of the spread prediction model during periods of extreme events, including two phases: the COVID-19 pandemic and the China-US trade conflicts. Finally, we check the validity of the benchmark model selection before.

### 4.2.1 Alternative training window periods

To avoid the influence of training window setting, we adjust the lengths of the rolling window and recursive window to 3 months and 12 months (M = 3, 12) to check the performance of various machine learning models in different window length settings. Panel A of Table 5 presents the results of MAE, RMSE, R², and $R^2_{OS}$ for each model trained using a 3-month rolling window. Consistent with the results in Table 2, the outcomes of the core evaluation criterion $R^2_{OS}$ indicate that when the rolling window length is set to 3 months, the predictive power of all machine learning models built based on traditional feature indicators is almost superior to that of all benchmark models. Furthermore, after incorporating corporate non-financial feature indicators, the out-of-sample predictive power of all models significantly improves. These results verify the robustness of our findings above.

**[Table 5 inserts here]**

Panel B of Table 5 shows the MAE, RMSE, R², and $R^2_{OS}$ results for each model trained using a 3-month recursive window. Consistent with the results in Table 3 and Panel A of Table 5, when the window length is set to 3 months (M = 3), the predictive power of all machine learning models constructed based on the recursive window and traditional feature indicators is already significantly better than that of the benchmark models. Moreover, after adding corporate non-financial feature indicators, the out-of-sample predictive power of all models is significantly enhanced, which once again, verifies the robustness of the conclusions.

Meanwhile, Panels C and D of Table 5 respectively present the MAE, RMSE, R², and $R^2_{OS}$ results for each model trained using a 12-month rolling window and a 12-month recursive window. Consistent with the previous results, the findings of the core evaluation criterion $R_{OS}^2$ demonstrate that when the rolling window length is set to 12 months, the predictive power of almost all machine learning models built based on traditional feature indicators is superior to that of all benchmark models. Additionally, after incorporating corporate non-financial feature indicators, the out-of-sample predictive power of all models is significantly improved. The results in Table 5 fully confirm the robustness of the conclusions in our study.

### 4.2.2 Predictability during extreme event periods

Furthermore, as extreme tail events significantly increase the overall credit risk in the market, we further test the performance of machine learning models in predicting credit spreads during extreme periods. Based on the sample period of this study, we select the China-US trade conflicts and the COVID-19 pandemic shock as extreme event periods, and adjust the sample periods to January 2018–December 2019 and January 2020–December 2022 respectively for the robustness tests.

**China-US trade conflicts.** Table 6 presents the predictive performance of various machine learning models with 6-month rolling window during the China-US trade conflicts period (2018-2019). Results of $R_{OS}^2$ show that, during the China-US trade conflicts, 3 out of the 7 machine learning models based on traditional feature indicators exhibit significantly better credit spread predictive performance than the traditional credit rating model, demonstrating the superiority of machine learning models. In particular, after incorporating corporate non-financial feature indicators, the predictive performance of all machine learning models was significantly improved, and the credit spread predictive performance of all machine learning models significantly surpassed that of the traditional rating model. This result indicates that even during periods of extreme shocks, the machine learning-driven credit spread prediction model constructed in our study still demonstrates superior performance. Specifically, the incorporation of corporate non-financial indicators significantly enhances the predictive power of machine learning models, further verifying the excellent predictive performance during periods of extreme events.

**[Table 6 inserts here]**

**COVID-19 pandemic shock.** Corresponding to Table 6, Table 7 presents the predictive performance of various machine learning models trained using a 6-month rolling window during the COVID-19 pandemic period (2020-2022)[7]. From the results of $R_{OS}^2$, during the COVID-19 pandemic, even the machine learning-driven credit spread prediction models built on traditional feature indicators all demonstrated better predictive performance than the traditional credit rating model. In particular, after the introduction of corporate non-financial feature indicators, the credit spread predictive performance of the machine learning models was further improved. Overall, the results in Table 7 all validate the conclusions of Table 6, confirming that our machine learning

---

[7] We selected the period from 2020 to 2022 as the COVID-19 pandemic period for the following reasons: In January 2020, after Chinese authorities confirmed the high transmissibility of the virus, they implemented strict measures and locked down Wuhan, marking China's entry into a special period of COVID-19 prevention and control. While at the end of 2022, China decided to fully lift COVID-19 prevention and control measures, which marked the end of the special period of the COVID-19 pandemic.

models exhibit excellent predictive performance during periods of extreme events.

**[Table 7 inserts here]**

### 4.2.3 Effectiveness of the benchmark model selection

In addition, we previously use the credit rating model as the benchmark model for credit spread prediction. However, credit rating model may not be a suitable benchmark with good predictive performance. To check this, we further evaluate the validity of the comparison benchmark by comparing the predictive performance of the credit rating model to other credit spread prediction models. Specifically, we use the traditional Altman Z-score model (Jiang et al., 2023) to replace the credit rating model for predicting bond credit spreads, thereby assessing the effectiveness of our benchmark model.

As shown in the results of Table 8, the predictive performance of the benchmark model (credit rating model) for credit spreads is significantly superior to that of the Altman Z-score model. This is reflected in the benchmark model having lower MAE, RMAE, and a higher R², while the $R^2_{OS}$ of the Altman Z-score model relative to the benchmark model is also negative. These results indicate that the market-based credit rating model is indeed superior to the traditional Altman Z-score model in predicting credit spreads, which also confirms the validity of our benchmark model selection.

**[Table 8 inserts here]**

## 5. Further analysis

### 5.1 Which indicators matter in credit spread prediction model

After verifying the superiority of machine learning models over traditional credit risk models, we use the RF model—with the best predictive performance—as the representative of machine learning models to further analyze the importance of various feature variables during the sample period, identify key predictive factors, and thereby reveal the internal decision-making logic of machine learning models in predicting credit spreads. To this end, we introduce the SHAP (SHapley Additive exPlanations) analytical framework. Based on the game theory-based Shapley value theory, SHAP explains the model by treating features as "game participants" and prediction results as "total payoff," and calculating the contribution of each feature (SHAP value). As an advance model interpretability technique, SHAP can penetrate the "black box" of complex models. It combines theoretical rigor (consistent with the axiom of fairness), global and local interpretability, universality across various models, and intuitive understandability, accurately attributing prediction results to each input feature indicator. Through calculating the positive/negative signs and magnitudes of SHAP values, we can respectively reveal the direction of impact (promoting or inhibiting) and the intensity of influence of specific features on credit spread predictions.

We conduct predictions using rolling windows and recursive windows respectively, and measure the importance of each feature variable through SHAP analysis. Specifically, by calculating the mean of SHAP values and the mean of absolute SHAP values for each indicator in the current validation set, we measure the average direction of impact and average magnitude of impact of the indicator at the monthly level. To construct a more robust global feature importance system spanning the entire sample period, we further aggregate the analysis results of all monthly windows. Finally, we determine their long-term direction of impact and overall contribution by calculating the average SHAP values and average absolute SHAP values of indicators at each level over the full sample period. The former serves as a core metric for judging the long-term positive/negative correlation between indicators and credit spreads, while the latter constitutes a key basis for measuring the overall predictive power of features. This analytical paradigm, which combines high predictive accuracy with strong interpretability, provides data-driven empirical insights for deeply understanding the formation mechanism of credit spreads in China.

Table 9 takes the RF model (the model with the best predictive performance) as an example and shows the top 20 indicators ranked by importance, as determined by average absolute SHAP values. Panel A presents results for models trained using rolling windows, from which we can find that when traditional feature indicators are used for prediction, 18 out of the top 20 indicators in terms of importance are corporate financial indicators, with only 1 macroeconomics indicator (regional GDP growth rate) and 1 bond-level variable (remaining maturity). This indicates that corporate financial features hold significant reference value for predicting bond credit spread, compared with macro-level features and bond-level features. Further analysis shows that the most important corporate financial indicator is book equity. The remaining corporate financial indicators, in order of importance, are: market value, total assets, return on assets, return on equity attributable to parent firm, price-to-cash flow ratio, return on equity (ROE), net profit margin excluding non-recurring gains and losses, ratio of total assets to current liabilities, cash ratio, ratio of advertising expenses to total assets, sales growth rate, operating net profit margin, total operating revenue, accounts receivable turnover, accounts receivable days, operating profit margin, and asset growth rate. From the distribution of important indicators, it can be seen that the top 20 indicators mainly cover multiple corporate financial dimensions, including profitability, firm size, financial coverage, liquidity, growth, R&D and innovation, operating efficiency, and valuation. This validates the rationality of the indicator set constructed in our study. In particular, indicators related to profitability and firm size exhibit the highest importance, with 6 and 4 indicators respectively ranking among the top 20. This is followed by the growth and operating efficiency dimensions, each with 2 indicators in the top 20. The liquidity, financial coverage, R&D and innovation, and valuation dimensions each have only 1 indicator in the top 20.

More importantly, after incorporating corporate non-financial indicators into the traditional feature indicators, **10 out of the top 20 indicators in the RF model (by importance) come from corporate non-financial indicators—7 of which rank among the top 10—while the number of financial indicators drops to 10. Further analysis reveals that the most important indicator in the RF model is the G-score** **of ESG rating (corporate governance).** The remaining indicators, in order of importance, are: nature of property rights, book equity, information disclosure evaluation, shareholding ratio of the largest shareholder, market value, shareholding ratio of senior executives, registered capital, audit report opinion, return on assets (ROA), total assets, ESG composite score, shareholding ratio of the board of directors, return on equity attributable to parent firm, price-to-cash flow ratio, total operating revenue, customer concentration, operating net profit margin, return on equity (ROE), and operating profit margin. All these indicators belong to the firm-level category. In addition, from the perspective of the importance distribution of non-financial indicators, all 5 non-financial feature dimensions we constructed have indicators ranking among top 20. Among these dimensions, the internal firm conditions and corporate governance performance are the most important, with 4 and 3 indicators respectively ranking in the top 20. They are followed by the dimensions from external firm environment, supply chain, and external support, each with 1 indicator in the top 20. These results not only verify the rationality of incorporating non-financial indicators, but also confirm the crucial role of firm-level indicators—especially non-financial indicators—in enhancing the credit spread predictive performance of machine learning models. More importantly, non-financial indicators are significantly more important than traditional feature indicators.

**[Table 9 inserts here]**

Next, from the results of models trained with recursive windows (presented in Panel B), when traditional feature indicators are used for prediction, the top 20 indicators by importance are still dominated by corporate financial indicators—specifically, 17 corporate financial indicators and 1 macroeconomic indicator (end-of-period M2) rank among the top 20. However, unlike models trained with rolling windows, this training method shows 2 bond-level indicators (plain vanilla bonds and issuance scale) enter the top 20. These results still demonstrate that under this training method, corporate financial indicators remain the most important predictive level. Further analysis shows that the most important indicator is still book equity. The other top-ranked indicators mainly cover multiple dimensions of corporate financial features (including firm size, profitability, liquidity, financial coverage, and cash flow), as well as macroeconomic indicators and bond-feature indicators. This validates the rationality of the indicator system constructed in our study. That said, there are slight differences in the importance of dimensions compared with models trained with rolling windows. Among these, indicators related to profitability and firm size (within the corporate financial dimension) remain the most prominent, with 7 and 4 indicators respectively ranking in the top 20. The leverage dimension follows, with 3 indicators on the list. For macroeconomic factors, the key indicator shifts to end-of-period M2, while for bond-level features, the key indicators shift to issuance scale and plain vanilla bonds.

Finally, **consistent with the results of models trained with rolling windows, panel B shows that after incorporating corporate non-financial indicators into the traditional feature indicators, 9 out of the top 20 indicators (by importance) in the**

**RF model are from corporate non-financial indicators—6 of which rank among the top 10—while the number of financial indicators drops to 9, the number of macroeconomic factor variables remains at 1, and the number of bond-level indicators decreases to 1.** Moreover, the most important indicator under recursive window training also comes from corporate non-financial indicators, which is nature of property rights. It is followed, in order of importance, by book equity, G-score in ESG rating, information disclosure evaluation, total assets, market value, shareholding ratio of the largest shareholder, registered capital, return on assets, ESG composite score, operating net profit margin, shareholding ratio of the board of directors, return on equity attributable to parent firm, end-of-period M2, shareholding ratio of senior executives, return on equity, plain vanilla bonds, institutional shareholding, cash flow to liability ratio, and total liabilities.

In addition, similar to the results of rolling window training, 3 out of the 5 corporate non-financial dimensions have indicators ranking top 20. Among these, indicators of the internal firm conditions and corporate governance dimension remain the most important, with 4 and 3 indicators respectively ranking in the top 20, and the external firm environment dimension has 2 indicators in the top 20. These results further verify the rationality of the non-financial indicator dimensions constructed in our study, and also further confirm the crucial role of firm-level indicators—especially non-financial indicators—in enhancing the credit spread predictive performance of machine learning models. After incorporating these indicators, the importance of other indicators decreases significantly. More importantly, corporate non-financial indicators exhibit more stable characteristics under different training methods, with little change in the importance of dimensions. This indicates that non-financial indicators can be combined with financial indicators, macro-level indicators, and bond-level indicators to demonstrate stable risk prediction performance.

## 5.2 Discussion on the role of non-financial indicators in credit spread prediction models

Table 2, Table3, and Table 9 all indicate a typical characteristic: non-financial indicators play a crucial role in improving the predictive performance of machine learning-driven credit spread prediction models. Not only the incorporation of these non-financial indicators significantly enhance the predictive performance of machine learning models, but these indicators also rank high in terms of importance. Among the top 20 indicators by importance, non-financial indicators account for 50% (using rolling window training) and 45% (using recursive window training). Moreover, among the top 10 most important predictors, 7 (using rolling window training) and 6 (using recursive window training) are from non-financial indicators, with their proportion exceeding 60%.

In particular, the innovatively introduced indicators related to a firm's internal conditions (such as nature of property rights, G-score, registered capital, and ESG composite score) and indicators related to corporate governance (such as the shareholding ratio of the largest shareholder, the shareholding ratio of senior executives, and the shareholding ratio of the board of directors) rank among the top most frequently. Additionally, the innovatively included indicators reflecting a firm's external environment—such as information disclosure evaluation indicators—and indicators reflecting supply chain conditions (e.g., customer concentration) also rank high.

These results fully confirm the effectiveness of incorporating non-financial indicators into our indicator set. By innovatively including non-financial indicators, our machine learning models could more effectively capture dynamic risk hidden in corporate operations, reveal potential risks such as corporate governance failures and intensified information asymmetry in advance, and address the lag defect of traditional macro indicators, financial indicators, and bond-level indicators.

## 5.3 Dissecting the economic mechanism of credit spread prediction model

An often mentioned drawback of machine learning models is their lack of interpretability. To this end, we refer to the method of Jiang et al. (2023) and use portfolio analysis to dig deeper into the drivers of the prediction performance. **In general, a rise in the credit risk of corporate bonds is logically accompanied by processes such as deteriorating operations, increasing leverage, short-termization of debt structure, and tighter corporate financing constraints.** In more severe cases, corporate financing pressure surges sharply, leading to the outbreak of debt default risks. We reveal the specific mechanism by which machine learning models predict corporate bond credit risks by examining whether such methods can better identify the processes closely associated with bond default risks.

Specifically, we first use the return on assets (ROA) as an indicator of corporate operating performance, the asset-liability ratio to measure the leverage level of firms, the ratio of current liabilities to total liabilities to measure the maturity structure of corporate debt, and the KZ index to measure the financing constraints of firms. Then, at the end of each month, we use the expected spread of each bond in the next month predicted by each model, divide the bonds into 5 equal groups in descending order of expected risk, and calculate the operating performance, leverage level, debt maturity structure, and financing constraints of the bond portfolio with the top 20% of predicted credit spreads (Group H) and the bond portfolio with the bottom 20% (Group L) respectively. Finally, we calculate the differences (H-L) in operating performance, leverage level, debt maturity structure, and financing constraints between the top 20% and bottom 20% bond portfolios, and conduct tests using the Newey-West t-statistic.

Table 10 reports the identification and prediction performance of various machine learning models regarding corporate operating deterioration, rising leverage, short-termization of debt structure, and tighter financing constraints. The empirical results show that all machine learning models can effectively identify significant differences between high-risk and low-risk companies in terms of operating performance, debt maturity structure, and financing constraints, and these differences (H-L) are all statistically significant at the 1% level. For firms with higher predicted credit spread risks (Group H) indicated by various models, their operating performance is worse, their debt is more short-termized, and their financing constraints are tighter.

At the same time, we find that none of the machine learning models can identify differences in leverage between high-risk and low-risk firms. In fact, firm flagged as high-risk by the models may even have lower leverage. Further analysis reveals that there is a strong linear relationship between the bond portfolios from the highest credit spread risk (Q5) to the lowest credit spread risk (Q1) identified by the machine learning models, and the trends of operating deterioration, short-termization of debt structure, and rising financing constraints. In contrast, there is a non-linear U-shaped relationship between these portfolios and leverage. Therefore, we believe this may be because the machine learning models distinguish between "good leverage" and "bad leverage" and capture a more profound logic than the simple linear relationship of "high leverage = high risk". Low-risk firms (Group L) are likely healthy firms with good investment opportunities, and they have the ability and willingness to increase liabilities to pursue development, so their leverage levels are not necessarily low. In contrast, high-risk firms (Group H) may be firms that have fallen into financial distress. Their credit ratings are poor, making it difficult for them to obtain loans, as banks and investors will avoid them. As a result, they cannot continue to increase leverage and even sell assets to repay debts, leading to stagnant or even declining leverage levels. In addition, the level of leverage is also closely related to the industry in which a firm operates, and leverage levels vary significantly across different industries (e.g., banking and manufacturing). The models may capture this industry heterogeneity and recognize that when assessing a firm's risk, profitability (ROA) and financing capacity (KZ index) are more fundamental and important determinants than leverage itself.

In summary, the results of Table 10 show that various machine learning models can effectively use the three core channels—operating deterioration, short-termization of debt, and financing constraints—to predict credit risks, revealing the economic mechanism behind various machine learning-driven credit spread prediction models. The "abnormal" result that the leverage mechanism is not significant seems to reveal that machine learning models can capture complex non-linear relationships, conduct high-dimensional modeling based on multi-dimensional indicators, thereby going beyond the simple accumulation of single financial indicators and achieving more accurate prediction of credit spreads.

**[Table 10 inserts here]**

## 6. Economic Value: Machine learning-driven credit spread prediction implied rating model

Previous analysis demonstrates that machine learning-driven credit spread prediction models—especially those incorporating corporate non-financial features—provide more accurate credit spread predictions than traditional credit rating models. In fact, credit risk also implicitly contains rating information. In recent bond market, implicit ratings are being increasingly widely used in credit rating practice. An implicit rating refers to the practice of rating bonds using their market prices (i.e., credit spreads). Inspired by this, machine learning-driven credit spread prediction models should be able to rate bonds based on the implicit ratings derived from predicted credit spreads, and address the issues of conflicts of interest, lagging updates, and insufficient differentiation faced by traditional rating agencies (Xia, 2014; Jiang et al., 2023, 2024). To this end, we attempt to further construct a machine learning-driven implicit credit rating model for bond credit spread prediction by using credit spread prediction indicator set proposed in previous sections, and test the effectiveness of this credit rating model.

**6.1 Construction of credit spread prediction implicit rating model**

We construct a credit rating model based on credit spread prediction, leveraging the machine learning methods and credit risk prediction indicator set proposed in previous sections. Specifically, first, we convert the actual credit spreads of bonds into credit ratings using the implicit ratings embedded in credit spreads. We sort the bonds in ascending order of credit spreads, assign bond credit ratings to them, and categorize these ratings into 10 classes (AAA, AA+, AA, A, BBB, BB, B, CCC, CC, C).

Next, we use the credit risk prediction indicator set proposed earlier to predict the next month's implicit credit ratings of bonds. We select two machine learning models that performed best in credit spread prediction—RF and XGBoost model—to predict the credit ratings of bonds (covering all industry sectors) in the entire dataset (refer to the $R_{OS}^2$ results in Tables 2 and 3). In addition, using the RF and XGBoost models as examples, we further develop 7 sets of industry-specific classification rating models for the seven major industrial sectors defined by China International Trust and Investment Corporation (CITIC), including the consumer sector, infrastructure & real estate sector, manufacturing sector, healthcare & wellness sector, cyclical sector, technology sector, and financial sector. Given the large proportion of data from the infrastructure & real estate sector, particularly the real estate industry within it, we also develop an industry-specific classification rating model for the real estate industry—a first-tier industry under the infrastructure & real estate sector.

Finally, following the conventional practice in machine learning, we randomly split the dataset into a training set, a validation set, and a test set at an 8:1:1 ratio. We determine the optimal parameter values of the model based on the criterion of maximizing the F1-score on the validation set, and use these parameters to predict credit ratings on the test set to obtain the model's prediction accuracy (Accuracy)— that is, the prediction accuracy on the test set. We evaluate the model performance using three widely accepted metrics in classification tasks: Accuracy, Recall Rate, and F1-Score.

Accuracy is the most intuitive performance metric for classification, and it measures the proportion of samples correctly predicted by the model out of the total number of samples. Recall Rate assesses the model's ability to identify all true positive samples. For a specific category, Recall Rate is defined as the proportion of samples correctly predicted as belonging to that category by the model relative to the total number of actual samples in that category. The F1-Score is the harmonic mean of Precision (the proportion of truly positive samples among those predicted as positive by the model) and Recall Rate. It aims to balance these two metrics simultaneously and provide a more balanced comprehensive evaluation. Compared with Accuracy, the F1-Score is more robust when dealing with imbalanced datasets.

In credit rating prediction, we neither want to misclassify a high-quality bond (requiring high Precision) nor miss a high-risk bond (requiring high Recall). The F1-Score can effectively balance these two needs. Therefore, we use the Weighted-Average F1-Score as the core basis for comparing the comprehensive performance of different models. In addition, we adopt K-fold cross-validation to effectively avoid the randomness flaw of the traditional "single training set/validation set split" and improve the stability of the model.

**6.2 Effectiveness of credit spread prediction implicit rating model**

Table 11 presents the effectiveness test results of the machine learning-driven predictive implicit rating model after K-fold cross-validation (where K=5). From the results in Panel A, it can be shown that the Accuracy Rates of machine learning models such as RF and XGBoost for credit rating prediction across all industries are 81.09% and 82.79% respectively, both exceeding 80%. Their Recall Rates are 81.09% and 82.79% respectively, and their F1-scores are 81.04% and 82.75% respectively. These machine learning models demonstrate excellent predictive performance in credit rating prediction across all industries.

**[Table 11 inserts here]**

From the industry-specific rating results driven by the RF model in Panel B, machine learning models also exhibit outstanding predictive performance in industry-specific and sector-specific rating tasks. The Accuracy, Recall Rate, and F1-Score for each industry all exceed 70%. Among them, the credit rating prediction accuracy is the highest in fields such as the financial sector, real estate sector, manufacturing sector, cyclical sector, and infrastructure & real estate sector, with F1-Scores of 88.21%, 83.97%, 83.06%, 81.96%, and 81.45% respectively. In fact, except for the healthcare & wellness sector, the implicit credit rating model based on credit spreads we proposed achieves a credit rating prediction accuracy (based on the more comprehensive F1-score) of over 80% for all other industries.

The results in Panel C indicate that the implicit credit rating model based on credit spreads driven by the XGBoost model also delivers excellent performance in industry-specific and sector-specific rating tasks. The Accuracy, Recall Rate, and F1-Score for each industry all exceed 75%. Among these, it performs best in rating the real estate industry, with an F1-Score of 85.50, followed by the financial sector, manufacturing sector, infrastructure & real estate sector, and cyclical sector, with F1-Scores of 84.57%, 83.55%, 83.22%, and 82.55% respectively.

Above all, the results in Table 11 indicate that the implicit credit rating prediction models based on credit spreads we constructed achieve a prediction accuracy (Accuracy, Recall Rate, and F1-Score) of over 70%—whether for the entire industry, or for specific industries and sectors. In particular, except for the healthcare & wellness sector, the prediction accuracy for all other industries exceeds 75%, demonstrating excellent credit rating prediction performance.

Next, we further examine the importance of each indicator in the machine learning-driven credit spread prediction implicit rating model. We also take the credit rating model (full-industry model) trained by the RF model as an example and calculate the top 20 indicators in terms of importance in the model. From the results in Table 12, the most important indicator is still a corporate non-financial indicator, namely registered capital, followed by nature of property rights, end-of-period M2, plain vanilla bonds, remaining maturity, book equity, G-score in ESG rating, information disclosure evaluation, shareholding ratio of the largest shareholder, fixed-asset investment in the tertiary industry, regional per capita tax revenue, firm age, total fixed investment, total assets to current liabilities, shareholding ratio of senior executives, issuance maturity, leading index, shareholding ratio of the board of directors, return on assets, and end-of-period value of M1.

**[Table 12 inserts here]**

Further analysis reveals that among the top 20 indicators in terms of importance, 8 are from corporate non-financial indicators (among which 5 rank among the top 10 in importance), 6 are from macroeconomic indicators, and 3 each are from financial indicators and bond-level indicators. Corporate non-financial indicators and macroeconomic indicators account for 40% and 30% respectively in terms of quantity, while financial indicators and bond-level indicators only account for 15% each. In particular, non-financial indicators related to a firm's internal conditions and corporate governance, as well as macroeconomic indicators related to monetary policy, rank the highest in importance, far exceeding corporate financial indicators. These results not only further verify the rationality of our introduction of non-financial indicators into the credit spread prediction and credit rating indicator set but also prove that non-financial indicators play a more important role in improving the performance of machine learning-driven credit rating models than traditional indicators such as macro-level indicators, corporate financial indicators, and bond-level indicators.

In summary, the results in Tables 11 and 12 indicate that both the full-industry machine learning models and the industry-specific and sector-specific machine learning sub-models exhibit excellent credit rating prediction performance, and among them, non-financial indicators play a more important role than traditional indicators in enhancing the performance of machine learning-driven credit rating models.

## 7. Conclusion

We select all corporate bonds, enterprise bonds, medium-term notes, and short-term financing bills issued by China's A-share listed firms from 2012 to 2023 as sample. By integrating corporate non-financial data with traditional data at the macro, corporate financial, and bond features levels, a bond credit spread prediction set containing 167 indicators is constructed. Based on 7 leading machine learning methods, corporate bond credit spread prediction models are built to empirically test the effectiveness of these machine learning models and the underlying economic mechanisms, while demonstrating the excellent performance of machine learning models in predicting bond credit ratings.

Our findings show that compared with traditional credit rating models, machine learning models exhibit better credit spread prediction effectiveness. Furthermore, compared with credit spread prediction indicator set with traditional data, non-financial indicators, built using corporate non-financial data could provide incremental predictive performance for machine learning models—the out-of-sample predictive performance is increased by more than double on average, significantly enhancing the ability of machine learning models to predict credit spreads. Further analysis also reveals that among machine learning models incorporating corporate non-financial indicators, the RF and XGBoost model achieve the best credit spread prediction performance, significantly outperforming other machine learning models. In addition, indicator importance analysis using the RF model as an example further shows that corporate financial and non-financial indicators provide more important spread prediction effectiveness than macro-level and bond-level indicators. In particular, corporate non-financial features such as corporate governance, nature of property rights, and information disclosure evaluation offer the most favorable predictive signals for machine learning-driven credit spread prediction models. After incorporating corporate non-financial indicators, 7 out of the top 10 indicators in terms of importance are non-financial indicators and the importance of macro-level indicators and corporate financial indicators decreases significantly and is notably lower than that of non-financial indicators. Moreover, when different window training methods are adopted, corporate non-financial indicators can be combined with macro indicators, corporate financial indicators, and bond-level indicators to exhibit more stable predictive effectiveness, whereas the importance of macro-level, corporate financial-level, and bond-level indicators is greatly affected by window training methods.

Analysis of the economic mechanism reveals that machine learning models can identify bonds of firms with high-risk characteristics, such as deteriorating operations, short-termized debt structures, and tighter financing constraints. There is a non-linear U-shaped relationship between leverage and corporate credit spreads, implying that various machine learning models primarily achieve accurate credit spread prediction by conducting high-dimensional modeling of multi-dimensional indicators to more effectively identify credit risk accumulation signals, such as corporate operational deterioration, short-termized debt structures, and a sharp increase in corporate credit financing constraints.

Finally, we find that machine learning methods and the credit spread prediction indicator set we constructed also show great potential in credit rating prediction. Both machine learning models for the entire industry and industry-specific/sector-specific machine learning sub-models demonstrate excellent credit rating prediction performance, with accuracy rates exceeding 70%. In particular, for industries other than the healthcare & wellness sector, the accuracy rate is as high as over 75%. In these machine learning-driven credit rating models, non-financial indicators also play a more important role than traditional characteristic indicators such as macro-level features, corporate financial features, and bond-level features.

Our study not only reveals the enhancing effect of machine learning on bond credit spread prediction and its underlying economic mechanisms, and demonstrates their great potential in credit rating prediction, but also shows that incorporating corporate non-financial data into machine learning models significantly improves the models' ability to predict credit spreads and credit ratings. Investors can proactively leverage corporate non-financial data to boost portfolio returns and strengthen risk management. Furthermore, while this study primarily focuses on predicting bond credit spreads and ratings, we could actually enable effective prediction of issuers' credit risk and rating status by aggregating bonds by issuer—providing a methodological foundation for assessing issuers' credit risk and ratings. In addition, we show the differences in the predictive performance of the industry-specific sub-rating models: for example, the financial, real estate, and manufacturing sectors perform excellently, while the healthcare & wellness sector performs relatively poorly. This indicates that the importance of indicators varies when predicting different industries, which provides room for fine-tuning and improvement in the development of industry-specific rating prediction models. Finally, our conclusions also show that corporate non-financial indicators—such as corporate governance structure and corporate ESG performance— are sound and stable indicators for credit spread prediction. This suggests that governments and regulatory authorities in emerging market countries should increase their focus on and supervision of corporate governance to mitigate risks in the bond market.

# Figure 1 Overall research framework

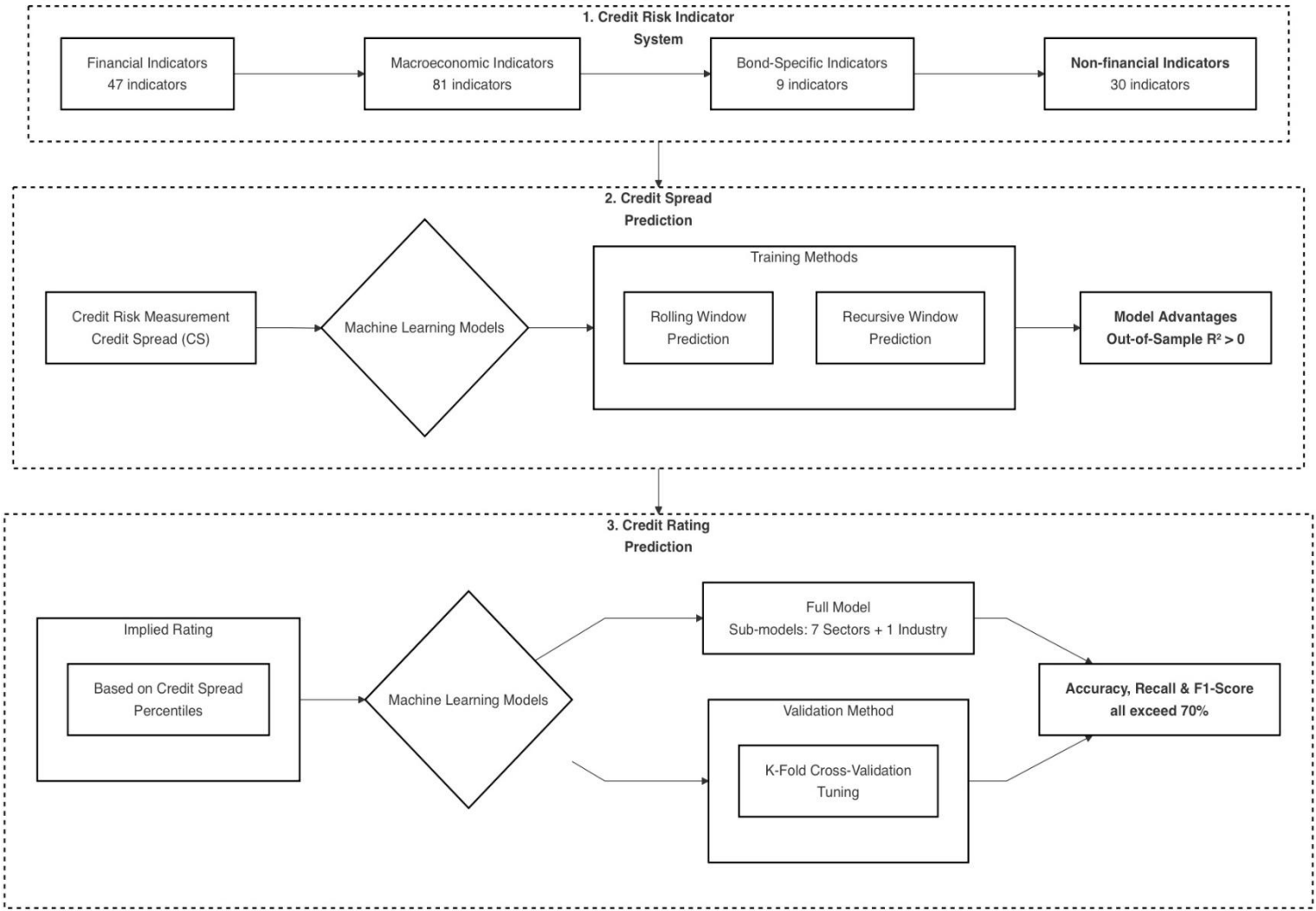

**Note:** This figure illustrates our overall research framework.

**Figure 2 Actual credit spreads vs. predicted values from the machine learning credit spread model for Vanke 22 MTN005 and 22 GN001 in 2023**

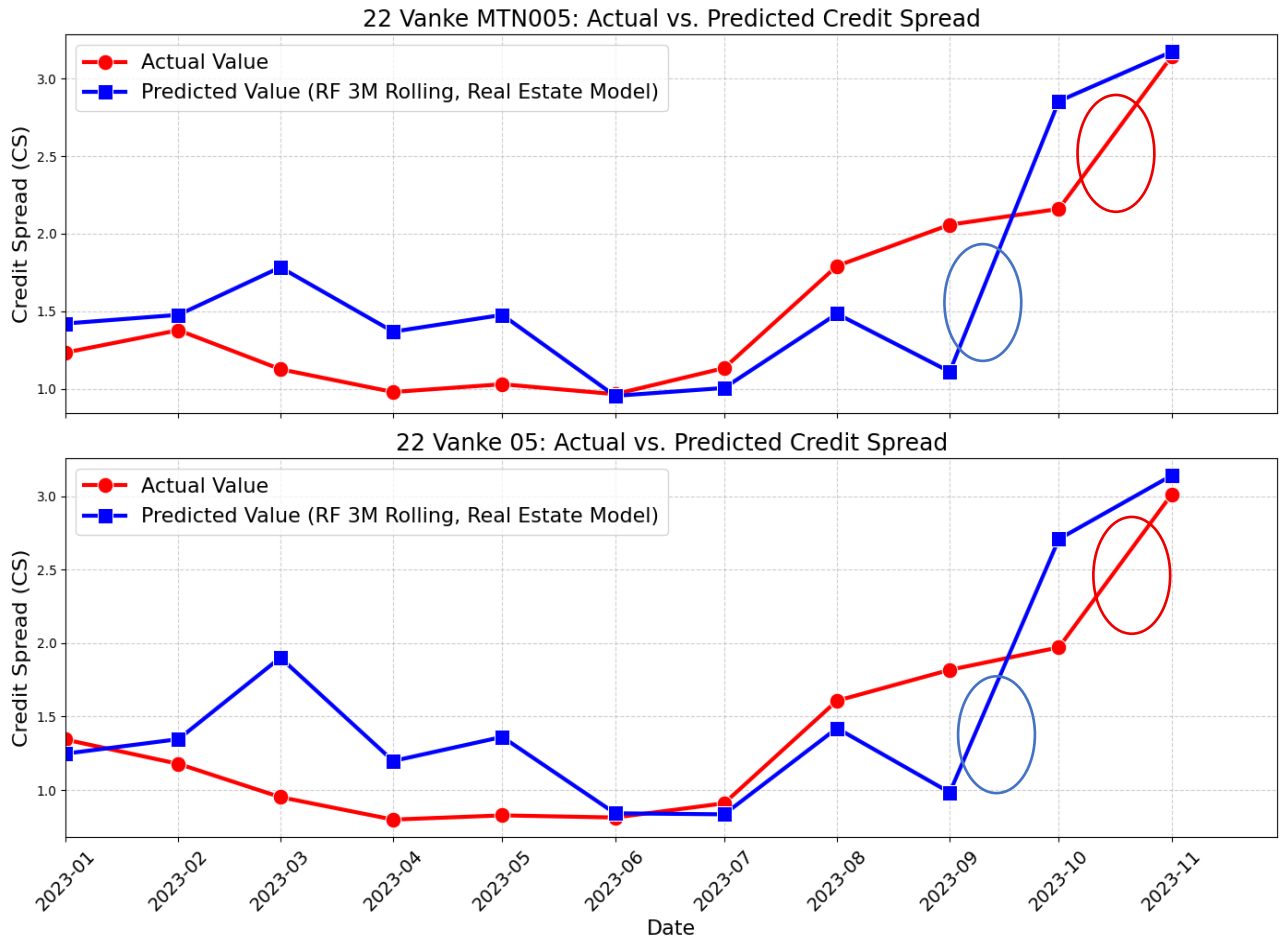

**Note:** This figure illustrates actual credit spreads vs. predicted values from the machine learning credit spread model for Vanke 22 MTN005 and 22 GN001 in 2023. The vertical axis represents credit spreads, and the horizontal axis represents the months of 2023. The red line stands for the actual credit spreads of bonds, while the blue line represents the predicted values from the credit spread prediction model (i.e., we predict the next month's credit spreads of bonds one month in advance). The predicted values are based on a real estate industry credit spread prediction model with using all credit risk prediction indicators and the Random Forest method. The monthly spread predictions lead all model indicators by one month and are derived from the model. For example, the predicted value for October 2023 is calculated and predicted at the end of September.

**Table 1 A brief summary of credit risk indicator set**

| Macro-level indicators | | Firm-level financial indicators | | Firm-level non-financial indicators | | Bond-level indicators | |
|---|---|---|---|---|---|---|---|
| Variables | Name | Variables | Name | Variables | Name | Variables | Name |
| M1 | Regional GDP Growth Rate | F1 | Return on Asset (ROA, TTM) | NF1 | Registered Capital | B1 | Issuance Maturity |
| M2 | Regional GDP Per Capita | F2 | Return on Equity (ROE, TTM) | NF2 | Firm Age | B2 | Remaining Maturity |
| M3 | Regional Per Capita Fiscal Revenue | F3 | Return on Equity Attributable to Parent Firm | NF3 | ESG Composite Score | B3 | Trading Venue |
| M4 | ······ | F4 | Gross Profit Margin | NF4 | ······ | B4 | Issuance Scale |
| M14 | Economic Policy Uncertainty Index | F5 | Net Profit Margin | NF10 | Overtime Work (Based on Nighttime Light Data) | B5 | Guarantee Status |
| M15 | ······ | F6 | ······ | NF11 | ······ | B6 | Plain Vanilla Bonds |
| M40 | Industrial Solid Waste Generation (Year-on-Year) | F22 | Current Ratio | NF14 | Proportion of Negative News | B7 | Callable Bond Only |
| M41 | ······ | F23 | ······ | NF15 | ······ | B8 | Puttable Bond Only |
| M81 | Pipeline Oil and Gas Transportation Turnover (Year-on-Year) | F47 | Asset-to-Market Ratio | NF30 | Abnormal Executive Departure | B9 | Other Types of Bonds |

**Note:** Detailed credit risk indicators are shown in Appendix B.

**Table 2 Out-of-sample predictability of machine learning models with rolling**

**window training**

| Model | Only traditional indicators (Macro level, firm financial level, and bond level indicators) | | | | Traditional indicators + Firm non-financial indicators | | | |
|---|---|---|---|---|---|---|---|---|
| | MAE | RMSE | $R^2$ | $R^2_{OS}$ | MAE | RMSE | $R^2$ | $R^2_{OS}$ |
| RF | 0.883 | 1.297 | 0.415 | 0.296 | 0.726 | 1.094 | 0.617 | 0.539 |
| AdaBoost | 1.049 | 1.448 | 0.278 | 0.131 | 0.873 | 1.248 | 0.506 | 0.406 |
| XGBoost | 0.876 | 1.321 | 0.374 | 0.247 | 0.687 | 1.081 | 0.624 | 0.547 |
| GBDT | 0.905 | 1.358 | 0.336 | 0.201 | 0.733 | 1.137 | 0.601 | 0.520 |
| LASSO | 1.102 | 1.593 | 0.173 | 0.005 | 0.994 | 1.425 | 0.372 | 0.244 |
| Ridge | 1.155 | 1.673 | 0.077 | -0.111 | 1.066 | 1.490 | 0.325 | 0.188 |
| Enet | 1.076 | 1.568 | 0.212 | 0.052 | 0.950 | 1.386 | 0.424 | 0.307 |
| Average | 1.007 | 1.465 | 0.266 | 0.117 | 0.861 | 1.266 | 0.496 | 0.393 |
| Benchmark | 1.141 | 1.686 | 0.167 | 0.000 | 1.141 | 1.686 | 0.167 | 0.000 |

**Note:** This table shows the out-of-sample predictive performance of the 7 machine learning models with rolling window. MAE, RMSE, $R^2$, and $R^2_{OS}$ represent predictive performance indicators, namely mean absolute error, root mean squared error, out-of-sample R-squared, and out-of-sample predictive R-squared, respectively. "Average" denotes the average results of the predictive performance indicators across the 7 machine learning models. "Benchmark" denotes the predictive performance of the benchmark model (i.e., the credit rating model).

**Table 3 Out-of-sample predictability of machine learning models with recursive window training**

| Model | Only traditional indicators (Macro level, firm financial level, and bond level indicators) | | | | Traditional indicators + Firm non-financial indicators | | | |
|---|---|---|---|---|---|---|---|---|
| | MAE | RMSE | $R^2$ | $R^2_{OS}$ | MAE | RMSE | $R^2$ | $R^2_{OS}$ |
| RF | 0.773 | 1.212 | 0.533 | 0.438 | 0.667 | 1.046 | **0.684** | **0.620** |
| AdaBoost | 1.188 | 1.634 | 0.132 | -0.045 | 1.036 | 1.428 | 0.397 | 0.275 |
| XGBoost | 0.775 | 1.194 | 0.543 | 0.449 | 0.663 | 1.022 | **0.699** | **0.638** |
| GBDT | 0.796 | 1.223 | 0.521 | 0.424 | 0.692 | 1.067 | 0.649 | 0.578 |
| LASSO | 1.093 | 1.616 | 0.216 | 0.056 | 0.971 | 1.448 | 0.393 | 0.269 |
| Ridge | 1.104 | 1.613 | 0.197 | 0.033 | 0.979 | 1.452 | 0.383 | 0.257 |
| Enet | 1.087 | 1.611 | 0.212 | 0.052 | 0.960 | 1.448 | 0.382 | 0.256 |
| Average | 0.974 | 1.443 | 0.336 | 0.201 | 0.852 | 1.273 | 0.513 | 0.413 |
| Benchmark | 1.141 | 1.686 | 0.167 | 0.000 | 1.141 | 1.686 | 0.167 | 0.000 |

**Note:** This table shows the out-of-sample predictive performance of the 7 machine learning models with recursive window. MAE, RMSE, $R^2$, and $R^2_{OS}$ represent predictive performance indicators, namely mean absolute error, root mean squared error, out-of-sample R-squared, and out-of-sample predictive R-squared, respectively. "Average" denotes the average results of the predictive performance indicators across the 7 machine learning models. "Benchmark" denotes the predictive performance results of the benchmark model (i.e., the credit rating model).

**Table 4 Diebold-Mariano test for the out-of-sample predictability of all models**

| Model | RF | AdaBoost | XGBoost | GBDT | LASSO | Ridge |
|---|---|---|---|---|---|---|
| AdaBoost | **4.75*** | | | | | |
| XGBoost | -0.82 | **-4.76*** | | | | |
| GBDT | **2.92*** | **-4.16*** | **3.42*** | | | |
| LASSO | **5.67*** | **3.44*** | **5.65*** | **5.16*** | | |
| Ridge | **5.41*** | **3.74*** | **5.46*** | **5.15*** | 1.74 | |
| ENet | **5.80*** | **3.13*** | **5.69*** | **5.29*** | **-2.91*** | **-2.46** |

**Note:** This table reports the pairwise Diebold-Mariano test t-statistics for comparing the out-of-sample credit spread prediction performance among seven models. A positive value in the table indicates that the prediction performance of the column model is superior to that of the row model. Bold font denotes that, for a single pairwise test, the difference in model performance is statistically significant at the 5% significance level or higher; while the asterisk (*) indicates that after accounting for 6 sets of multiple comparisons via our conservative Bonferroni adjustment method, the difference remains statistically significant at the 5% level.

**Table 5 Out-of-sample predictability of models trained with 3-month and 12-month windows**

Panel A：Models trained based on 3-month rolling window

| Model | Only traditional indicators (Macro level, firm financial level, and bond level indicators) | | | | Traditional indicators + Firm non-financial indicators | | | |
|---|---|---|---|---|---|---|---|---|
| | MAE | RMSE | $R^2$ | $R^2_{OS}$ | MAE | RMSE | $R^2$ | $R^2_{OS}$ |
| RF | 0.768 | 1.114 | 0.517 | 0.419 | 0.656 | 0.972 | **0.679** | **0.614** |
| AdaBoost | 0.921 | 1.285 | 0.376 | 0.250 | 0.747 | 1.066 | 0.631 | 0.557 |
| XGBoost | 0.789 | 1.178 | 0.385 | 0.261 | 0.594 | 0.940 | **0.710** | **0.652** |
| GBDT | 0.789 | 1.170 | 0.432 | 0.317 | 0.628 | 0.968 | 0.684 | 0.620 |
| LASSO | 1.077 | 1.546 | 0.183 | 0.018 | 0.939 | 1.331 | 0.433 | 0.319 |
| Ridge | 1.469 | 1.976 | -0.610 | -0.935 | 1.251 | 1.695 | -0.096 | -0.318 |
| Enet | 1.003 | 1.470 | 0.277 | 0.130 | 0.876 | 1.271 | 0.482 | 0.377 |
| Average | 0.974 | 1.391 | 0.223 | 0.066 | 0.813 | 1.178 | 0.503 | 0.403 |
| Benchmark | 1.141 | 1.686 | 0.167 | 0.000 | 1.141 | 1.686 | 0.167 | 0.000 |

Panel B：Models trained based on 3-month recursive window

| Model | Only traditional indicators (Macro level, firm financial level, and bond level indicators) | | | | Traditional indicators + Firm non-financial indicators | | | |
|---|---|---|---|---|---|---|---|---|
| | MAE | RMSE | $R^2$ | $R^2_{OS}$ | MAE | RMSE | $R^2$ | $R^2_{OS}$ |
| RF | 0.632 | 1.006 | 0.641 | 0.568 | 0.555 | 0.878 | **0.763** | **0.715** |
| AdaBoost | 1.131 | 1.536 | 0.188 | 0.024 | 0.992 | 1.349 | 0.431 | 0.316 |
| XGBoost | 0.640 | 1.000 | 0.639 | 0.566 | 0.553 | 0.858 | **0.773** | **0.727** |
| GBDT | 0.662 | 1.029 | 0.626 | 0.551 | 0.588 | 0.911 | 0.728 | 0.673 |
| LASSO | 1.044 | 1.558 | 0.235 | 0.080 | 0.926 | 1.392 | 0.411 | 0.292 |
| Ridge | 1.108 | 1.613 | 0.207 | 0.047 | 0.936 | 1.410 | 0.403 | 0.282 |
| Enet | 1.048 | 1.563 | 0.226 | 0.069 | 0.923 | 1.399 | 0.398 | 0.276 |
| Average | 0.895 | 1.329 | 0.394 | 0.272 | 0.782 | 1.171 | 0.558 | 0.469 |
| Benchmark | 1.141 | 1.686 | 0.167 | 0.000 | 1.141 | 1.686 | 0.167 | 0.000 |

Panel C: Models trained based on 12-month rolling window

| Model | Only traditional indicators (Macro level, firm financial level, and bond level indicators) | | | | Traditional indicators + Firm non-financial indicators | | | |
|---|---|---|---|---|---|---|---|---|
| | MAE | RMSE | $R^2$ | $R^2_{OS}$ | MAE | RMSE | $R^2$ | $R^2_{OS}$ |
| RF | 0.897 | 1.386 | 0.417 | 0.296 | 0.758 | 1.194 | 0.605 | 0.523 |
| AdaBoost | 1.104 | 1.550 | 0.256 | 0.102 | 0.962 | 1.349 | 0.478 | 0.369 |
| XGBoost | 0.879 | 1.376 | 0.429 | 0.311 | 0.752 | 1.190 | 0.605 | 0.523 |
| GBDT | 0.906 | 1.425 | 0.395 | 0.269 | 0.791 | 1.254 | 0.549 | 0.455 |
| LASSO | 1.099 | 1.621 | 0.227 | 0.067 | 0.999 | 1.465 | 0.387 | 0.259 |
| Ridge | 1.128 | 1.636 | 0.204 | 0.039 | 1.006 | 1.460 | 0.390 | 0.264 |

| | MAE | RMSE | $R^2$ | $R^2_{OS}$ | MAE | RMSE | $R^2$ | $R^2_{OS}$ |
|---|---|---|---|---|---|---|---|---|
| Enet | 1.090 | 1.616 | 0.233 | 0.074 | 0.977 | 1.448 | 0.411 | 0.289 |
| Average | 1.015 | 1.516 | 0.309 | 0.165 | 0.892 | 1.337 | 0.489 | 0.383 |
| Benchmark | 1.141 | 1.686 | 0.167 | 0.000 | 1.141 | 1.686 | 0.167 | 0.000 |

Panel D: Models trained based on 12-month recursive window

| Model | Only traditional indicators (Macro level, firm financial level, and bond level indicators) | | | | Traditional indicators + Firm non-financial indicators | | | |
|---|---|---|---|---|---|---|---|---|
| | MAE | RMSE | $R^2$ | $R^2_{OS}$ | MAE | RMSE | $R^2$ | $R^2_{OS}$ |
| RF | 0.889 | 1.370 | 0.421 | 0.301 | 0.776 | 1.213 | 0.583 | 0.497 |
| AdaBoost | 1.246 | 1.716 | 0.088 | -0.101 | 1.084 | 1.511 | 0.341 | 0.204 |
| XGBoost | 0.880 | 1.353 | 0.435 | 0.318 | 0.777 | 1.197 | 0.586 | 0.500 |
| GBDT | 0.897 | 1.364 | 0.437 | 0.320 | 0.787 | 1.212 | 0.578 | 0.491 |
| LASSO | 1.142 | 1.674 | 0.187 | 0.018 | 1.008 | 1.500 | 0.368 | 0.237 |
| Ridge | 1.191 | 1.702 | 0.135 | -0.044 | 1.008 | 1.496 | 0.362 | 0.230 |
| Enet | 1.128 | 1.659 | 0.191 | 0.024 | 0.992 | 1.496 | 0.361 | 0.229 |
| Average | 1.053 | 1.548 | 0.271 | 0.119 | 0.919 | 1.375 | 0.454 | 0.341 |
| Benchmark | 1.141 | 1.686 | 0.167 | 0.000 | 1.141 | 1.686 | 0.167 | 0.000 |

**Note:** This table shows the out-of-sample predictive performance of the 7 machine learning models with 3-month or 12-month training window. MAE, RMSE, $R^2$, and $R^2_{OS}$ represent predictive performance indicators, namely mean absolute error, root mean squared error, out-of-sample R-squared, and out-of-sample predictive R-squared, respectively. "Average" denotes the average results of the predictive performance indicators across the 7 machine learning models. "Benchmark" denotes the predictive performance results of the benchmark model (i.e., the credit rating model).

**Table 6 Out-of-sample predictability of the models during the China-U.S. trade conflict (2018-2019)**

| Model | Only traditional indicators (Macro level, firm financial level, and bond level indicators) | | | | Traditional indicators + Firm non-financial indicators | | | |
|---|---|---|---|---|---|---|---|---|
| | MAE | RMSE | $R^2$ | $R^2_{OS}$ | MAE | RMSE | $R^2$ | $R^2_{OS}$ |
| RF | 1.007 | 1.471 | 0.401 | 0.223 | 0.815 | 1.254 | 0.607 | 0.490 |
| AdaBoost | 1.252 | 1.667 | 0.181 | -0.062 | 1.055 | 1.448 | 0.446 | 0.282 |
| XGBoost | 1.036 | 1.509 | 0.343 | 0.148 | 0.786 | 1.249 | 0.591 | 0.469 |
| GBDT | 1.090 | 1.572 | 0.293 | 0.084 | 0.864 | 1.332 | 0.557 | 0.426 |
| LASSO | 1.250 | 1.755 | 0.177 | -0.067 | 1.103 | 1.559 | 0.343 | 0.148 |
| Ridge | 1.358 | 1.897 | -0.177 | -0.526 | 1.158 | 1.628 | 0.230 | 0.001 |
| Enet | 1.246 | 1.752 | 0.224 | -0.006 | 1.067 | 1.534 | 0.414 | 0.240 |
| Average | 1.177 | 1.660 | 0.206 | -0.030 | 0.978 | 1.429 | 0.455 | 0.294 |
| Benchmark | 1.154 | 1.845 | 0.229 | 0.000 | 1.154 | 1.845 | 0.229 | 0.000 |

**Note:** This table shows the out-of-sample predictive performance of the 7 machine learning models with rolling window during the period of Sino-US trade conflict. MAE, RMSE, $R^2$, and $R^2_{OS}$ represent predictive performance indicators, namely mean absolute error, root mean squared error, out-of-sample R-squared, and out-of-sample predictive R-squared, respectively. "Average" denotes the average results of the predictive performance indicators across the 7 machine learning models. "Benchmark" denotes the predictive performance results of the benchmark model (i.e., the credit rating model).

**Table 7 Out-of-sample predictability of the models during COVID-19 pandemic**

**(2020-2022)**

| Model | Only traditional indicators (Macro level, firm financial level, and bond level indicators) | | | | Traditional indicators + Firm non-financial indicators | | | |
|---|---|---|---|---|---|---|---|---|
| | MAE | RMSE | $R^2$ | $R^2_{OS}$ | MAE | RMSE | $R^2$ | $R^2_{OS}$ |
| RF | 1.120 | 1.727 | 0.408 | 0.293 | 0.830 | 1.344 | 0.608 | 0.532 |
| AdaBoost | 1.386 | 1.960 | 0.278 | 0.139 | 1.036 | 1.573 | 0.505 | 0.410 |
| XGBoost | 1.086 | 1.740 | 0.373 | 0.252 | 0.750 | 0.535 | 0.623 | 0.550 |
| GBDT | 1.112 | 1.793 | 0.336 | 0.208 | 0.810 | 1.390 | 0.600 | 0.522 |
| LASSO | 1.417 | 2.113 | 0.188 | 0.031 | 1.220 | 1.795 | 0.418 | 0.306 |
| Ridge | 1.456 | 2.132 | 0.167 | 0.006 | 1.247 | 1.800 | 0.401 | 0.286 |
| Enet | 1.383 | 2.084 | 0.223 | 0.072 | 1.160 | 1.759 | 0.447 | 0.341 |
| Average | 1.280 | 1.936 | 0.282 | 0.143 | 1.008 | 1.456 | 0.515 | 0.421 |
| Benchmark | 1.312 | 2.252 | 0.162 | 0.000 | 1.312 | 2.252 | 0.162 | 0.000 |

**Note:** This table shows the out-of-sample predictive performance of the 7 machine learning models with rolling window during the period of COVID-19 pandemic. MAE, RMSE, $R^2$, and $R^2_{OS}$ represent predictive performance indicators, namely mean absolute error, root mean squared error, out-of-sample R-squared, and out-of-sample predictive R-squared, respectively. "Average" denotes the average results of the predictive performance indicators across the 7 machine learning models. "Benchmark" denotes the predictive performance results of the benchmark model (i.e., the credit rating model).

**Table 8 Test for effectiveness of benchmark model selection**

| Model | Altman Z | Benchmark model (credit rating model) |
|---|---|---|
| MAE | 1.304 | 1.312 |
| RMAE | 1.930 | 2.252 |
| $R^2$ | -0.058 | 0.162 |
| $R^2_{OS}$ | -0.270 | 0.000 |

**Note:** This table compares the predictive performance of the Altman Z model and the benchmark model (credit rating model) proposed in this paper for credit spreads. MAE, RMSE, $R^2$, and $R^2_{OS}$ represent predictive performance indicators, namely mean absolute error, root mean squared error, out-of-sample R-squared, and out-of-sample predictive R-squared, respectively. $R^2_{OS}$ is calculated based on the prediction results of credit rating model.

**Table 9 Variable importance analysis of machine learning model (RF model)**

Panel A: Models trained based on rolling window

| Rank | Only traditional indicators (Macro level, firm financial level, and bond level indicators) | | Traditional indicators + Firm non-financial indicators | |
| --- | --- | --- | --- | --- |
| | Name | Contribution degree | Name | Contribution degree |
| 1 | Book Equity: F18 | 0.192 | G-Score: NF6 | 0.194 |
| 2 | Market Value: F16 | 0.084 | Nature of Property Rights: NF7 | 0.175 |
| 3 | Total Assets: F15 | 0.076 | Book Equity: F18 | 0.108 |
| 4 | Return on Assets (ROA): F1 | 0.065 | Information Disclosure Evaluation: NF8 | 0.075 |
| 5 | Return on Equity Attributable to Parent Firm：F3 | 0.062 | Shareholding Ratio of the Largest Shareholder: NF23 | 0.066 |
| 6 | Price-to-Cash Flow Ratio: F46 | 0.058 | Market Value: F16 | 0.057 |
| 7 | Return on Equity (ROE): F2 | 0.052 | Shareholding Ratio of Senior Executives: NF24 | 0.054 |
| 8 | Net Profit Margin Excluding Non-Recurring Gains and Losses: F9 | 0.050 | Registered Capital: NF1 | 0.053 |
| 9 | Ratio of Total Assets to Current Liabilities: F21 | 0.049 | Audit Report Opinion: NF21 | 0.051 |
| 10 | Cash Ratio: F24 | 0.047 | Return on Assets (ROA): F1 | 0.050 |
| 11 | Ratio of Advertising Expenses to Total Assets: F33 | 0.046 | Total Assets: F15 | 0.050 |
| 12 | Sales Growth Rate: F28 | 0.045 | ESG Composite Score: NF3 | 0.049 |
| 13 | Operating Net Profit Margin: F5 | 0.045 | Shareholding Ratio of the Board of Directors: NF25 | 0.048 |
| 14 | Total Operating Revenue: F17 | 0.041 | Return on Equity Attributable to Parent firm：F3 | 0.042 |
| 15 | Accounts Receivable Turnover: F38 | 0.040 | Price-to-Cash Flow Ratio: F46 | 0.041 |
| 16 | Regional GDP Growth Rate: M1 | 0.039 | Total Operating Revenue: F17 | 0.038 |
| 17 | Accounts Receivable Turnover Days: F39 | 0.039 | Customer Concentration: NF15 | 0.037 |
| 18 | Operating Profit Margin: F6 | 0.038 | Operating Net Profit Margin: F5 | 0.036 |
| 19 | Asset Growth Rate: F29 | 0.037 | Return on Equity (ROE): F2 | 0.034 |
| 20 | Remaining Maturity: B2 | 0.035 | Operating Profit Margin: F6 | 0.032 |

Panel B: Models trained based on recursive window

| Rank | Only traditional indicators (Macro level, firm financial level, and bond level indicators) | | Traditional indicators + Firm non-financial indicators | |
| --- | --- | --- | --- | --- |
| | Name | Contribution degree | Name | Contribution degree |
| 1 | Book Equity: F18 | 0.328 | Nature of Property Rights: NF7 | 0.304 |
| 2 | Total Assets: F15 | 0.117 | Book Equity: F18 | 0.183 |
| 3 | Market Value: F16 | 0.105 | G-Score: NF6 | 0.179 |
| 4 | Ratio of Total Assets to Current Liabilities: F21 | 0.078 | Information Disclosure Evaluation: NF8 | 0.114 |
| 5 | Return on Assets (ROA): F1 | 0.077 | Total Assets: F15 | 0.080 |
| 6 | Operating Net Profit Margin: F5 | 0.052 | Market Value: F16 | 0.075 |
| 7 | Return on Equity Attributable to Parent Firm: F3 | 0.050 | Shareholding Ratio of the Largest Shareholder: NF23 | 0.055 |
| 8 | End-of-Period Value of M2: M20 | 0.049 | Registered Capital: NF1 | 0.053 |
| 9 | Total Liabilities: F19 | 0.049 | Return on Assets (ROA): F1 | 0.052 |
| 10 | Return on Equity (ROE): F2 | 0.042 | ESG Composite Score: NF3 | 0.051 |
| 11 | Debt-to-Asset Ratio: F11 | 0.040 | Operating Net Profit Margin: F5 | 0.042 |
| 12 | Long-Term Debt Ratio: F12 | 0.040 | Shareholding Ratio of the Board of Directors: NF25 | 0.039 |
| 13 | Net Profit Margin Excluding Non-Recurring Gains and Losses: F9 | 0.039 | Return on Equity Attributable to Parent Firm: F3 | 0.038 |
| 14 | Operating Profit Margin: F6 | 0.039 | End-of-Period Value of M2: M20 | 0.037 |
| 15 | Debt-to-Equity Ratio: F13 | 0.039 | Shareholding Ratio of Senior Executives: NF24 | 0.037 |
| 16 | Plain Vanilla Bonds: B6 | 0.037 | Return on Equity (ROE): F2 | 0.036 |
| 17 | Cash Ratio: F24 | 0.037 | Plain Vanilla Bonds: B6 | 0.035 |
| 18 | Issuance Scale: B4 | 0.035 | Institutional Shareholding: NF9 | 0.033 |
| 19 | Cash Flow to Liability Ratio: F37 | 0.035 | Cash Flow to Liability Ratio: F37 | 0.032 |
| 20 | Net Profit Margin: F8 | 0.034 | Total Liabilities: F19 | 0.031 |

**Note:** This table presents the results of variable importance analysis for the credit spread prediction model using the SHAP method (SHapley Additive exPlanations). To construct a more robust global feature importance system across the entire sample period, we measure the variable contribution by calculating the monthly average of the absolute SHAP values of indicators at all levels throughout the full sample period. For the detailed of each predictive indicator, please refer to Appendix B.

**Table 10 Economic mechanism analysis of credit spread prediction model**

| Economic mechanism | RF | AdaBoost | XGBoost | GBDT | LASSO | Ridge | Enet |
|---|---|---|---|---|---|---|---|
| Operation deterioration： | | | | | | | |
| Q1 (L) | 0.0377 | 0.0360 | 0.0377 | 0.0365 | 0.0414 | 0.0403 | 0.0412 |
| Q2 | 0.0378 | 0.0376 | 0.0365 | 0.0371 | 0.0354 | 0.0354 | 0.0357 |
| Q3 | 0.0345 | 0.0343 | 0.0333 | 0.0331 | 0.0316 | 0.0318 | 0.0325 |
| Q4 | 0.0273 | 0.0285 | 0.0277 | 0.0280 | 0.0274 | 0.0286 | 0.0276 |
| Q5 (H) | 0.0092 | 0.0097 | 0.0112 | 0.0118 | 0.0105 | 0.0104 | 0.0094 |
| H-L | **-0.0285** | **-0.0263** | **-0.0265** | **-0.0247** | **-0.0309** | **-0.0299** | **-0.0318** |
| t-value | -8.2742 | -8.3436 | -8.5037 | -8.6165 | -10.4724 | -10.5471 | -10.3472 |
| Leverage rising： | | | | | | | |
| Q1 (L) | 0.6348 | 0.6519 | 0.6288 | 0.6225 | 0.6116 | 0.6116 | 0.6161 |
| Q2 | 0.5983 | 0.5955 | 0.6001 | 0.6021 | 0.6099 | 0.6084 | 0.6068 |
| Q3 | 0.6045 | 0.5928 | 0.6059 | 0.6119 | 0.6154 | 0.6158 | 0.6123 |
| Q4 | 0.6087 | 0.6082 | 0.6132 | 0.6138 | 0.6187 | 0.6186 | 0.6205 |
| Q5 (H) | 0.6387 | 0.6305 | 0.6371 | 0.6346 | 0.6298 | 0.6308 | 0.6290 |
| H-L | 0.0039 | **-0.0214** | 0.0083 | 0.0121 | 0.0182 | 0.0193 | 0.0129 |
| t-value | 0.3111 | -2.4186 | 0.7298 | 1.1280 | 1.4239 | 1.6563 | 0.9338 |
| Short-termization of debt structure： | | | | | | | |
| Q1 (L) | 0.6541 | 0.6747 | 0.6618 | 0.6643 | 0.6569 | 0.6589 | 0.6551 |
| Q2 | 0.6772 | 0.6728 | 0.6739 | 0.6732 | 0.6802 | 0.6758 | 0.6776 |
| Q3 | 0.7002 | 0.6825 | 0.6935 | 0.6934 | 0.6936 | 0.6941 | 0.6928 |
| Q4 | 0.7094 | 0.7053 | 0.7131 | 0.7084 | 0.7087 | 0.7145 | 0.7161 |
| Q5 (H) | 0.7288 | 0.7365 | 0.7272 | 0.7301 | 0.7295 | 0.7263 | 0.7280 |
| H-L | **0.0747** | **0.0617** | **0.0654** | **0.0658** | **0.0726** | **0.0674** | **0.0729** |
| t-value | 4.3846 | 3.5869 | 4.4232 | 4.0683 | 5.0837 | 4.4899 | 4.4384 |
| Financing constraints increasing： | | | | | | | |
| Q1 (L) | 2.0578 | 2.1568 | 2.0043 | 2.0124 | 1.8088 | 1.8374 | 1.8092 |
| Q2 | 1.8611 | 1.8502 | 1.9171 | 1.9059 | 1.9968 | 2.0141 | 2.0059 |
| Q3 | 2.0233 | 1.9501 | 2.0449 | 2.0601 | 2.0850 | 2.0886 | 2.0594 |
| Q4 | 2.0825 | 1.9986 | 2.0823 | 2.0771 | 2.1531 | 2.1003 | 2.1437 |
| Q5 (H) | 2.4858 | 2.5090 | 2.4647 | 2.4580 | 2.4741 | 2.4748 | 2.4984 |
| H-L | **0.4280** | **0.3522** | **0.4604** | **0.4456** | **0.6653** | **0.6374** | **0.6893** |
| t-value | 4.0067 | 4.1672 | 4.7302 | 4.4604 | 7.1932 | 6.7844 | 7.5060 |

**Note:** This table presents the mechanism analysis results of the credit spread prediction model using the portfolio analysis method. At the end of each month, we evenly divide bonds into five groups based on the spread prediction results of the credit spread prediction model, sorted from low to high: Q1 represents the 20% of bonds with the lowest credit spreads (indicating lower credit risk), while Q5 represents the 20% with the highest credit spreads (indicating higher credit risk). We then use ROA (Return on Assets), asset-liability ratio, current liabilities to total liabilities, and KZ index to measure a firm's operating performance, leverage level, debt maturity structure, and financing constraints, respectively, and calculate the operating performance, leverage level, debt maturity structure, and financing constraints of each Q1-Q5 portfolio in the following month. Finally, we

compute the differences in operating performance, leverage level, debt maturity structure, and financing constraints between the Q5 portfolio (top 20%) and Q1 portfolio (bottom 20%)—denoted as the H-L portfolio (High-Low)—and conduct statistical tests using Newey-West t-statistics with the optimal lag length. H-L values displayed in bold black font are statistically significant at the 1% level.

**Table 11 Results of machine learning-driven implicit credit rating prediction**

**model based on credit spreads**

| Panel A: All industry credit rating model | | | |
|---|---|---|---|
| Model | Accuracy | Recall rate | F1 score |
| RF | 81.09% | 81.09% | 81.04% |
| XGBoost | 82.79% | 82.79% | 82.75% |
| Panel B: RF model-driven industry-specific credit rating model | | | |
| Industry | Accuracy | Recall rate | F1 value |
| Infrastructure and Real Estate Sector | 81.46% | 81.46% | 81.45% |
| — Real Estate Industry | 83.99% | 83.99% | 83.97% |
| Consumer Sector | 80.75% | 80.75% | 80.61% |
| Manufacturing Sector | 83.14% | 83.14% | 83.06% |
| Healthcare & Wellness Sector | 73.91% | 73.91% | 74.02% |
| Cyclical Sector | 82.11% | 82.11% | 81.96% |
| Technology Sector | 81.19% | 81.19% | 81.13% |
| Financial Sector | 87.50% | 87.50% | 88.21% |
| Panel C: XGBoost model-driven industry-specific credit rating model | | | |
| Industry | Accuracy | Recall rate | F1 score |
| Infrastructure and Real Estate Sector | 83.26% | 83.26% | 83.22% |
| — Real Estate Industry | 85.54% | 85.54% | 85.50% |
| Consumer Sector | 78.26% | 78.26% | 78.08% |
| Manufacturing Sector | 83.71% | 83.71% | 83.55% |
| Healthcare & Wellness Sector | 79.50% | 79.50% | 79.14% |
| Cyclical Sector | 82.72% | 82.72% | 82.55% |
| Technology Sector | 81.96% | 81.96% | 81.89% |
| Financial Sector | 84.38% | 84.38% | 84.57% |

**Note:** This table presents the predictive performance of the credit spread prediction-based implicit rating models driven by RF and XGBoost models. Specifically, Panel A shows the industry-wide credit spread prediction-based implicit rating models built with the RF and XGBoost methods. Panels B and C respectively present the industry-specific credit spread prediction-based implicit rating models constructed using RF and XGBoost models. For industry classification, we adopt the seven major industrial sectors defined by China International Trust and Investment Corporation (CITIC), including the consumer sector, infrastructure & real estate sector, manufacturing sector, healthcare & wellness sector, cyclical sector, technology sector, and financial sector. We also develop an industry-specific classification rating model for the real estate industry—a first-tier industry under the infrastructure & real estate sector.

**Table 12 Variable importance analysis of machine learning-driven (RF model)**

**implicit credit rating prediction model based on credit spreads**

| Rank | Name | Contribution degree |
|------|------|---------------------|
| 1 | Registered Capital: NF1 | 0.023 |
| 2 | Nature of Property Rights: NF7 | 0.020 |
| 3 | End-of-Period Value of M2: M20 | 0.015 |
| 4 | Plain Vanilla Bonds: B6 | 0.012 |
| 5 | Remaining Maturity: B2 | 0.010 |
| 6 | Book Equity: F18 | 0.009 |
| 7 | G-Score: NF6 | 0.007 |
| 8 | Information Disclosure Evaluation: NF8 | 0.007 |
| 9 | Shareholding Ratio of the Largest Shareholder: NF23 | 0.006 |
| 10 | Fixed Asset Investment in the Tertiary Industry: M57 | 0.006 |
| 11 | Regional Per Capita Tax Revenue: M4 | 0.005 |
| 12 | Firm Age: NF2 | 0.004 |
| 13 | Total Fixed Investment: M53 | 0.004 |
| 14 | Total Assets to Current Liabilities: F21 | 0.004 |
| 15 | Shareholding Ratio of Senior Executives: NF24 | 0.004 |
| 16 | Issuance Maturity: B1 | 0.003 |
| 17 | Leading Index: M13 | 0.003 |
| 18 | Shareholding Ratio of the Board of Directors: NF25 | 0.003 |
| 19 | Return on Assets (ROA): F1 | 0.003 |
| 20 | End-of-Period Value of M1: M22 | 0.003 |

**Note:** This table presents the results of variable importance analysis for RF model-driven credit spread prediction-based implicit rating using the SHAP method (SHapley Additive exPlanations). To construct a more robust global feature importance system across the entire sample period, we measure the variable contribution by calculating the monthly average of the absolute SHAP values of indicators at all levels throughout the full sample period. For the detailed of each predictive indicator, please refer to Appendix B.

**Appendix A Introduction to other machine learning models mentioned in manuscript**

**AdaBoost regression (AdaBoost)** improves the accuracy of credit risk prediction by iteratively optimizing the weighted combination of weak predictors. Its core logic is as follows: During the first training iteration, equal weights are assigned to all samples to fit and obtain an initial weak regressor; after calculating the prediction residuals, the weights of samples with large residuals are increased, and a new weak regressor is generated through retraining; this process is repeated until N weak regressors are constructed. Finally, the integrated prediction result is output by weighted summation of each regressor using an exponential loss function. This method can dynamically focus on the difficulties of risk identification. In credit spread prediction, it strengthens the learning of default-sensitive samples through iteration, enabling the prediction error to gradually converge as the number of weak regressors increases.

**Gradient Boosting Decision Tree (GBDT)** is an iterative decision tree model composed of multiple decision trees. It integrates the predictions of all trees to generate the final prediction. The core of this model lies in that each tree learns the residuals of all previous trees. To eliminate these residuals, the model builds a new model in the direction of the gradient where the residuals decrease. In the GBDT model, each new tree is constructed to reduce the residuals of the previous model along the gradient direction.

**LASSO regression (LASSO)** achieves feature selection and parameter shrinkage through L1 regularization, which adapts to the dimensionality reduction needs of the credit risk indicator system. When constructing a credit risk model, in the face of multi-dimensional indicators such as financial, macroeconomic, and non-financial ones, LASSO introduces a regularization penalty term to shrink the coefficients of unimportant features to zero, thereby automatically selecting core risk factors.

**Ridge regression (Ridge)** is a regression method that introduces an L2 regularization term based on ordinary least squares (OLS) regression. Its purpose is to alleviate the problem of multicollinearity and improve the generalization ability of the model. By adding the sum of squares of regression coefficients as a penalty term to the loss function, this method sacrifices a certain degree of fitting accuracy in exchange for more stable and robust regression coefficients. Specifically, Ridge regression minimizes the weighted sum of the residual sum of squares (RSS) and the L2 norm. The strength of regularization is controlled by the hyperparameter $\lambda$: the larger the $\lambda$, the stronger the suppression of model complexity, and the greater the degree of coefficient shrinkage. The final coefficients can be obtained through an analytical solution or numerical optimization methods (such as gradient descent), and they are usually smoother and more stable than those obtained by ordinary least squares.

**Elastic Net (Enet)** is a linear regression method that combines L1 and L2 regularization. It integrates the variable selection capability of LASSO regression and the stability of Ridge regression, making it particularly suitable for high-dimensional feature data with strong multicollinearity. Its loss function includes two penalty terms: the first is the sum of absolute values of coefficients (L1), which is used for sparse modeling; the second is the sum of squared coefficients (L2), which is used to control multicollinearity. The model adjusts two hyperparameters: λ controls the overall regularization strength, and adjusts the weight ratio between L1 and L2. When α approaches 1, the model is more similar to LASSO; when α approaches 0, it is closer to Ridge. Elastic Net is usually optimized using the coordinate descent method, which can simultaneously achieve variable selection and coefficient shrinkage, and exhibits better stability and predictive performance in practice.

# Appendix B Description of the indicator set for credit spread prediction

| Dimension | Symbol | Definition | Calculation Method (if applicable) | Data Source |
|---|---|---|---|---|
| **Macro-level predictive indicators** | | | | |
| Regional environmental indicators | M1 | Regional GDP growth rate | (Current period regional GDP-previous period regional GDP) / previous period regional GDP | Choice database |
| | M2 | Regional GDP per capita | Per capita regional gross domestic product (yuan/person) | |
| | M3 | Regional per capita fiscal revenue | Regional general budgetary revenue / Year-end permanent population | |
| | M4 | Regional per capita tax revenue | Regional fiscal tax revenue / Year-end permanent population | |
| | M5 | Regional per capita fiscal expenditure | Regional general budgetary expenditure / Year-end permanent population | |
| Macro-economic indicators | M6 | Industrial added value year-on-year growth rate | | Choice database |
| | M7 | CPI year-on-year growth rate | | |
| | M8 | PPI year-on-year growth rate | | |
| | M9 | RPI year-on-year growth rate | | |
| | M10 | GDP | | |
| | M11 | Per capita GDP | | |
| | M12 | Unemployment rate | | |
| | M13 | Leading index | | |
| | M14 | Economic policy uncertainty index | | |

| | M15 | Year-on-year growth rate of power generation | | |
|---|---|---|---|---|
| | M16 | Year-on-year growth rate of cement production | | |
| | M17 | Year-on-year change in the SHANGHAI SHENZHEN 300 INDEX | | |
| | M18 | Year-on-year change in the CSI Aggregate Bond Index | | |
| | M19 | Consumer confidence index (CCI) | | Wind database |
| | M20 | End-of-Period value of M2 | | |
| | M21 | Year-on-year growth rate of M2 | | |
| | M22 | End-of-Period value of M1 | | |
| | M23 | Year-on-year growth rate of M1 | | Choice database |
| | M24 | End-of-Period value of M0 | | |
| | M25 | Year-on-year growth rate of M0 | | |
| | M26 | Year-on-year growth rate of Aggregate | | National Bureau of Statistics |

| | | Financing to the Real Economy | | |
|---|---|---|---|---|
| | M27 | Year-on-year growth rate of new yuan-denominated loans | | |
| Resource (Year-on-Year) | M28 | Gold reserves | | |
| | M29 | Silver ore reserves | | |
| | M30 | Silver production | | |
| | M31 | Gold production | | |
| | M32 | Crude steel production | | |
| | M33 | Pig iron production | | |
| | M34 | Proven oil reserves | Choice database | |
| Agriculture (Year-on-Year) | M35 | Rice sown area | | |
| | M36 | Wheat sown area | | |
| | M37 | Corn sown area | | |
| | M38 | Sugar crops sown area | | |
| | M39 | Cotton sown area | | |
| Environment (Year-on-Year) | M40 | Industrial solid waste generation | | |
| | M41 | Industrial solid waste comprehensive utilization amount | | |
| | M42 | Forest area | | |
| | M43 | Forest coverage rate | CSMAR database | |
| | M44 | Nature reserve area | | |
| | M45 | Wetland area | | |
| | M46 | Total water resources | | |
| | M47 | Per capita water resources | | |

| | M48 | Total water supply | | |
|---|---|---|---|---|
| | M49 | Total water consumption | | |
| | M50 | Total wastewater discharge | | |
| | M51 | Sulfur dioxide emissions | | |
| | M52 | Soot emissions | | |
| Fixed investment (Year-on-Year) | M53 | Total fixed investment | | |
| | M54 | Urban fixed asset investment | | |
| | M55 | Fixed asset investment in the primary industry | | |
| | M56 | Fixed asset investment in the secondary industry | | |
| | M57 | Fixed asset investment in the tertiary industry | | |
| | M58 | Floor area of buildings under construction | | |
| | M59 | Floor area of completed buildings | | |
| Passenger Traffic (Year-on-Year | M60 | Total passenger volume | | |
| | M61 | Railway passenger volume | | |
| | M62 | Civil aviation passenger volume | | |
| | M63 | Highway passenger volume | | |

| | | | | | |
|---|---|---|---|---|---|
| | M64 | Waterway passenger volume | | | |
| Passenger Turnover (Year-on-Year) | M65 | Total passenger turnover volume | | | |
| | M66 | Civil aviation passenger turnover volume | | | |
| | M67 | Total railway passenger turnover volume | | | |
| | M68 | Highway passenger turnover volume | | | |
| | M69 | Waterway passenger turnover volume | | | |
| Freight Traffic (Year-on-Year) | M70 | Total freight traffic | | | |
| | M71 | Railway freight volume | | | |
| | M72 | Civil aviation cargo volume | | | |
| | M73 | Highway freight volume | | | |
| | M74 | Waterway freight volume | | | |
| | M75 | Pipeline oil & gas transportation volume | | | |
| Freight Turnover (Year-on-Year) | M76 | Total cargo turnover | | | |
| | M77 | Railway cargo turnover | | | |
| | M78 | Civil aviation cargo turnover | | | |

| | M79 | Highway cargo turnover | | |
|---|---|---|---|---|
| | M80 | Waterway cargo turnover | | |
| | M81 | Pipeline oil and gas transportation turnover | | |
| **Corporate financial-level prediction indicators** | | | | |
| Profitability | F1 | Return on asset (ROA, TTM) | (Net Profit TTM) / Average Total Assets, Average Total Assets = (Ending Balance of Total Assets + Year-ago Same-period Ending Balance of Total Assets) / 2 | CSMAR database |
| | F2 | Return on equity (ROE, TTM) | (Net Profit TTM) / Average Shareholders' Equity, Average Shareholders' Equity = (Ending Balance of Shareholders' Equity + Year-ago Same-period Ending Balance of Shareholders' Equity) / 2 | |
| | F3 | Return on equity attributable to parent firm | (Net Profit Attributable to Owners of the Parent Company) TTM / Average Equity Attributable to Owners of the Parent Firm, Average Equity Attributable to Owners of the Parent Firm= (Ending Balance of Equity Attributable to Owners of the Parent Firm+ Year-ago Same-period Ending Balance of Equity Attributable to Owners of the Parent Firm) / 2 | |
| | F4 | Gross profit margin | (Operating Revenue - Operating Cost) TTM / (Operating Revenue) TTM | |
| | F5 | Net profit margin | (Net Profit) TTM / (Operating Revenue) TTM | |
| | F6 | Operating profit margin | Operating Profit / Total Operating Revenue | |
| | F7 | Ebit operating profit margin TTM | (Net Profit + Income Tax Expense + Financial Expenses) TTM / (Operating Revenue) TTM | |
| | F8 | Net profit margin | Net Profit / Operating Revenue | |
| | F9 | Net profit margin excluding non-recurring gains and losses | Net Profit Attributable to Shareholders of Listed Firms (Excluding Non-Recurring Gains and Losses) / Total Operating Revenue | |
| | F10 | Return on net operating assets | Operating Revenue / Operating Assets | |
| Leverage | F11 | Debt-to-Asset Ratio | Total Liabilities / Total Assets. If the numerator is empty, the result is empty | CSMAR database |
| | F12 | Long-term debt ratio | Long-term Liabilities / Total Assets | |
| | F13 | Debt-to-Equity ratio | Long-term Liabilities / Net Assets | |
| | F14 | Long-term debt to equity ratio | Total Long-term Liabilities / Total Equity | |
| Firm size | F15 | Total asset | Total of Each Item of Assets | CSMAR database |
| | F16 | Market value | Total Market Value of Individual Stock | |
| | F17 | Total Operating Revenue | The sum of all revenues in the course of a firm's business operations. Total Operating Revenue = Operating Revenue + Net Interest Income + Earned Premiums + Net Commission and Fee Income + Other Operating Income | |
| | F18 | Book Equity | Total of Each Item of Shareholders' Equity | |

| | F19 | Total Liabilities | Total of Each Item of Liabilities | |
|---|---|---|---|---|
| Financial coverage | F20 | Cash flow interest coverage ratio | Net Cash Flow from Operating Activities / Financial Expenses | CSMAR database |
| | F21 | Total assets to current liabilities | Total Assets / Total Current Liabilities | |
| Liquidity | F22 | Current Ratio | Current Assets / Current Liabilities | CSMAR database |
| | F23 | Quick asset ratio | (Current Assets - Inventory) / Current Liabilities | |
| | F24 | Cash ratio | Ending Balance of Cash and Cash Equivalents / Current Liabilities | |
| | F25 | Ratio of operational capital to total assets | (Current Assets - Current Liabilities) / Total Assets | |
| Asset structure | F26 | Proportion of fixed assets to total assets | Fixed Assets / Total Assets | CSMAR database |
| | F27 | Book-to-Market ratio | Book Value / Market Value | |
| Growth | F28 | Sales growth rate | (Current Period Sales - Previous Period Sales) / Previous Period Sales | CSMAR database |
| | F29 | Asset growth rate | (Current Period Total Assets - Previous Period Total Assets) / Previous Period Total Assets | |
| | F30 | Inventory growth rate | (Current Period Net Inventory - Previous Period Net Inventory) / Previous Period Net Inventory | |
| R&D and innovation | F31 | Ratio of R&D expenditures to total assets | R&D Expenditures / Total Assets | CSMAR database |
| | F32 | Ratio of R&D expenditures to sales | R&D Expenditures / Total Operating Revenue | |
| | F33 | Ratio of advertising expenses to total assets | Sales Expenses / Total Assets | |
| Cash flow | F34 | Cash flow from operating activities | The difference between cash inflows from operating activities and cash outflows from operating activities | CSMAR database |
| | F35 | Cash productivity | （MV-TPA）/L<br>MV: Market value of the firm's equity (number of shares multiplied by stock price) plus the book value of the firm's debt based on COMPUSTAT. TPA: Book value of assets. L: Cash and short-term investments.） | |
| | F36 | Cash flow to stock price Ratio | Cash Flow from Operating Activities / Stock Price | |
| | F37 | Cash flow to liability ratio | Cash Flow from Operating Activities / Total Liabilities | |
| Operational capacity | F38 | Accounts receivable turnover TTM | (Operating Revenue) TTM / Average Balance of (Net Accounts Receivable), Average Balance of Net Accounts Receivable = (Current Period Ending Value of Net Accounts Receivable + Year-ago Same-period Ending Value of Net Accounts Receivable) / 2;; | CSMAR database |
| | F39 | Accounts receivable turnover days TTM | 365 / Accounts Receivable Turnover TTM | |
| | F40 | Inventory turnover TTM | (Operating Cost) TTM / Average Balance of Net Inventory, Average Balance of Net Inventory = (Current Period Ending Value of Net Inventory + Year-ago Same-period Ending Value of Net Inventory) / 2 | |
| | F41 | Inventory turnover days TTM | 365 / Inventory Turnover TTM | |

| | F42 | Ratio of sales volume to cash flow | Total Operating Revenue / Cash Flow from Operating Activities | |
| | F43 | Ratio of sales volume to inventory | Total Operating Revenue / Net Inventory | |
| Valuation | F44 | Price-to-earnings ratio (PE) | Current Period Value of Today's Closing Price / (Net Profit TTM / Current Period Ending Value of Paid-in Capital) | CSMAR database |
| | F45 | Price-to-sales ratio (PS) | Current Period Value of Today's Closing Price / (Total Operating Revenue TTM / Current Period Ending Value of Paid-in Capital) | |
| | F46 | Price-to-cash flow ratio (PCF) | Current Period Value of Today's Closing Price / (Net Cash Flow from Operating Activities TTM / Current Period Ending Value of Paid-in Capital) | |
| | F47 | Asset-to-market ratio | Total Assets / Market Value | |
| **Corporate non-financial-level prediction indicators** | | | | |
| Internal firm conditions | NF1 | Registered capital | | Choice database |
| | NF2 | Firm age | Firm Establishment Time (in months), Frequency: Monthly | CSMAR database |
| | NF3 | ESG composite score | Sino-Securities ESG Rating, ESG Composite Score | Sino-Securities ESG Rating database |
| | NF4 | E-Score | Sino-Securities ESG Rating, E Score | |
| | NF5 | S-Score | Sino-Securities ESG Rating, S Score | |
| | NF6 | G-Score | Sino-Securities ESG Rating, G Score | |
| | NF7 | Nature of property right | Dummy Variable: 1 for state-owned enterprises (SOEs), 0 otherwise | CSMAR database |
| External firm environment | NF8 | Information disclosure evaluation | Evaluation Result: 1=Excellent; 2=Good; 3=Qualified; 4=Unqualified | CSMAR database |
| | NF9 | Institutional shareholding | Total Shares Held by Institutional Investors / Total Shares of Listed Company | |
| | NF10 | Overtime work | Dummy Variable: 1 if there is overtime work in the municipal government of the company's location in the current month, 0 otherwise | Black Marble database |
| | NF11 | Total number of News | Total Number of News Items | CNRDS database |
| | NF12 | Proportion of positive News | Positive News / Total Number of News Items | |
| | NF13 | Proportion of neutral News | Neutral News / Total Number of News Items | |
| | NF14 | Proportion of negative News | Negative News / Total Number of News Items | |
| Supply chain | NF15 | Customer concentration | Sales to Top 5 Customers / Annual Total Sales | CSMAR database |
| | NF16 | Supplier concentration | Procurements from Top 5 Suppliers / Annual Total Procurement Amount | |
| External support | NF17 | Government subsidies | Government Subsidies / Annual Net Profit (Earnings Per Share × Total Number of Shares) | CSMAR database |
| | NF18 | Political connection | Whether the Chairman or General Manager Has Political Background; Frequency: Annual | Wind database |
| | NF19 | Degree of political connection | Institutional Level of Position Held by the Chairman orCEO, Frequency: Monthly | |
| | NF20 | Delayed disclosure | Dummy Variable: 1 if the disclosure time of financial reports exceeds one month after the current quarter, 0 otherwise | CSMAR database |
| | NF21 | Audit report opinion | Dummy Variable: 1 if the audit report opinion is "Standard Unqualified Opinion", 0 otherwise | |
| Corporate governance | NF22 | Dual position of chairman and CEO | Dummy Variable: 1 if the Chairman and CEO are the same person (concurrent position), 0 otherwise | CSMAR database |
| | NF23 | Shareholding Ratio of the Largest Shareholder | Shareholding Ratio of the Largest Shareholder | |
| | NF24 | Shareholding ratio of senior executives | Shareholding Quantity of Directors, Supervisors and Senior Management / Total Number of Shares | |
| | NF25 | Shareholding ratio of the board of directors | Shareholding Quantity of the Board of Directors / Total Number of Shares | CNRDS database |
| | NF26 | Total number of directors | Total Number of Directors | |
| | NF27 | Total number of supervisors | Total Number of Supervisors | |
| | NF28 | Proportion of independent directors | Number of Independent Directors / Total Number of Directors | |
| | NF29 | Total remuneration of the top three senior executives | Total Remuneration of the Top Three Senior Executives | |
| | NF30 | Abnormal executive departure | Dummy Variable: 1 if the reason for departure is resignation, dismissal, health reasons, personal reasons, involvement in cases, or other reasons, 0 otherwise | CSMAR database |
| **Bond-level prediction indicators** | | | | |
| Bond feature | B1 | Issuance maturity | | Choice database |
| | B2 | Remaining maturity | | |
| | B3 | Trading venue | | |

| | B4 | Issuance scale | | |
|---|---|---|---|---|
| | B5 | Guarantee status | | |
| | B6 | Plain vanilla bonds | Dummy variable: takes a value of 1 for bonds without put, call, or other provisions; otherwise, 0. | CSMAR database |
| | B7 | Callable bond only | Dummy variable: takes a value of 1 for bonds with only call provisions; otherwise, 0. | |
| | B8 | Puttable bond only | Dummy variable: takes a value of 1 for bonds with only put provisions; otherwise, it takes a value of 0. | |
| | B9 | Other types of bonds | Dummy variable: takes a value of 1 for bonds not in the B6-B8 category; otherwise, 0. | |

## Appendix C Descriptive statistical results of the credit spread prediction

### indicator set

| Variable | N | Mean | Min | Max | Median | Standard Error |
|---|---|---|---|---|---|---|
| Credit spread | 45283 | 2.082 | 0.423 | 11.442 | 1.285 | 2.226 |
| Credit spread _benchmark | 45283 | 2.098 | 1.581 | 11.482 | 1.581 | 0.886 |
| M1 | 45283 | 0.077 | 0.005 | 0.215 | 0.080 | 0.042 |
| M2 | 45283 | 103064.390 | 27303.000 | 189988.000 | 88521.000 | 45087.338 |
| M3 | 45283 | 13328.182 | 2637.626 | 30739.394 | 10218.044 | 8598.386 |
| M4 | 45283 | 5786.654 | 3899.298 | 6386.438 | 5940.004 | 495.725 |
| M5 | 45283 | 18849.066 | 5797.272 | 34199.405 | 14385.848 | 9202.385 |
| M6 | 45283 | 5.359 | -1.100 | 14.100 | 5.300 | 2.125 |
| M7 | 45283 | 1.762 | -0.500 | 5.200 | 1.800 | 1.232 |
| M8 | 45283 | 1.536 | -5.923 | 12.900 | -0.500 | 5.062 |
| M9 | 45283 | 1.713 | -1.100 | 4.207 | 1.700 | 1.088 |
| M10 | 45283 | 0.445 | 0.183 | 0.700 | 0.500 | 0.187 |
| M11 | 45283 | 0.424 | 0.122 | 0.705 | 0.482 | 0.205 |
| M12 | 45283 | 0.363 | 0.302 | 0.458 | 0.338 | 0.055 |
| M13 | 45283 | 99.223 | 92.790 | 105.010 | 99.020 | 3.676 |
| M14 | 45283 | 301.464 | 60.206 | 649.073 | 290.408 | 116.449 |
| M15 | 45283 | 3.949 | -4.600 | 17.400 | 4.000 | 3.777 |
| M16 | 45283 | -0.707 | -18.900 | 28.666 | -0.200 | 9.062 |
| M17 | 45283 | 4.134 | -33.541 | 106.594 | 1.818 | 20.462 |
| M18 | 45283 | 4.682 | -1.938 | 10.250 | 4.857 | 2.130 |
| M19 | 45283 | 112.820 | 86.300 | 130.700 | 121.200 | 16.153 |
| M20 | 45283 | 2.179 | 1.033 | 2.897 | 2.172 | 0.445 |
| M21 | 45283 | 10.169 | 8.000 | 15.800 | 10.300 | 1.704 |
| M22 | 45283 | 587684.380 | 307648.420 | 679588.350 | 616113.170 | 87720.679 |
| M23 | 45283 | 6.263 | 0.000 | 25.300 | 4.900 | 4.635 |
| M24 | 45283 | 84564.294 | 54063.910 | 110225.180 | 84314.530 | 13820.904 |
| M25 | 45283 | 8.144 | -13.800 | 19.400 | 7.700 | 4.305 |
| M26 | 45283 | 0.090 | -0.140 | 0.469 | 0.021 | 0.192 |
| M27 | 45283 | 0.093 | -0.005 | 0.186 | 0.084 | 0.063 |
| M28 | 45283 | 0.128 | -0.015 | 0.538 | 0.043 | 0.189 |
| M29 | 45283 | 0.088 | -0.020 | 0.417 | 0.027 | 0.152 |
| M30 | 45283 | 0.007 | -0.236 | 0.471 | -0.006 | 0.128 |
| M31 | 45283 | -0.022 | -0.099 | 0.131 | -0.052 | 0.074 |
| M32 | 45283 | 0.035 | -0.028 | 0.123 | 0.067 | 0.046 |
| M33 | 45283 | 0.035 | -0.031 | 0.100 | 0.037 | 0.048 |
| M34 | 45283 | 0.020 | 0.001 | 0.032 | 0.019 | 0.006 |
| M35 | 45283 | -0.006 | -0.018 | 0.013 | -0.005 | 0.011 |
| M36 | 45283 | -0.007 | -0.022 | 0.008 | -0.008 | 0.010 |
| M37 | 45283 | 0.005 | -0.040 | 0.064 | -0.006 | 0.029 |
| M38 | 45283 | -0.018 | -0.095 | 0.050 | -0.008 | 0.038 |
| M39 | 45283 | -0.023 | -0.153 | 0.050 | -0.009 | 0.044 |
| M40 | 45283 | 0.015 | -0.055 | 0.251 | 0.004 | 0.057 |
| M41 | 45283 | -0.003 | -0.074 | 0.120 | -0.008 | 0.033 |
| M42 | 45283 | 0.010 | 0.000 | 0.063 | 0.000 | 0.022 |
| M43 | 45283 | 0.010 | 0.000 | 0.061 | 0.000 | 0.022 |
| M44 | 45283 | 0.000 | -0.023 | 0.005 | 0.001 | 0.004 |
| M45 | 45283 | 0.008 | 0.000 | 0.393 | 0.000 | 0.056 |
| M46 | 45283 | 0.001 | -0.114 | 0.270 | -0.045 | 0.083 |
| M47 | 45283 | -0.002 | -0.119 | 0.261 | -0.050 | 0.083 |
| M48 | 45283 | -0.002 | -0.035 | 0.018 | 0.001 | 0.018 |
| M49 | 45283 | -0.002 | -0.035 | 0.018 | 0.001 | 0.018 |
| M50 | 45283 | 0.022 | -0.033 | 0.039 | 0.027 | 0.015 |
| M51 | 45283 | -0.190 | -0.407 | -0.034 | -0.137 | 0.105 |
| M52 | 45283 | -0.114 | -0.343 | 0.362 | -0.116 | 0.094 |
| M53 | 45283 | 0.003 | -0.131 | 0.203 | 0.048 | 0.085 |
| M54 | 45283 | 0.006 | -0.132 | 0.207 | 0.059 | 0.087 |
| M55 | 45283 | 0.014 | -0.578 | 0.271 | 0.098 | 0.264 |
| M56 | 45283 | 0.001 | -0.440 | 0.226 | 0.017 | 0.193 |
| M57 | 45283 | 0.025 | -0.202 | 0.218 | 0.065 | 0.103 |
| M58 | 45283 | 0.039 | -0.072 | 0.161 | 0.052 | 0.053 |
| M59 | 45283 | -0.017 | -0.150 | 0.112 | -0.044 | 0.083 |
| M60 | 45283 | -0.174 | -0.451 | 0.079 | -0.141 | 0.174 |
| M61 | 45283 | -0.046 | -0.398 | 0.185 | 0.084 | 0.235 |
| M62 | 45283 | -0.068 | -0.429 | 0.130 | 0.079 | 0.228 |
| M63 | 45283 | -0.209 | -0.479 | 0.082 | -0.262 | 0.168 |
| M64 | 45283 | -0.108 | -0.450 | 0.117 | -0.026 | 0.198 |
| M65 | 45283 | -0.112 | -0.455 | 0.077 | 0.026 | 0.210 |
| M66 | 45283 | -0.076 | -0.461 | 0.150 | 0.093 | 0.251 |
| M67 | 45283 | -0.072 | -0.438 | 0.157 | 0.040 | 0.222 |
| M68 | 45283 | -0.200 | -0.476 | 0.102 | -0.218 | 0.174 |
| M69 | 45283 | -0.144 | -0.589 | 0.088 | 0.004 | 0.241 |
| M70 | 45283 | 0.022 | -0.085 | 0.121 | 0.003 | 0.070 |
| M71 | 45283 | 0.050 | -0.119 | 0.107 | 0.049 | 0.046 |
| M72 | 45283 | -0.009 | -0.170 | 0.082 | 0.020 | 0.089 |
| M73 | 45283 | 0.014 | -0.132 | 0.142 | -0.003 | 0.092 |
| M74 | 45283 | 0.054 | 0.019 | 0.220 | 0.052 | 0.032 |
| M75 | 45283 | 0.018 | -0.103 | 0.131 | 0.016 | 0.070 |
| M76 | 45283 | 0.034 | -0.033 | 0.107 | 0.037 | 0.045 |
| M87 | 45283 | 0.047 | -0.137 | 0.133 | 0.069 | 0.053 |
| M88 | 45283 | 0.024 | -0.087 | 0.158 | 0.003 | 0.092 |
| M89 | 45283 | 0.017 | -0.163 | 0.159 | 0.009 | 0.100 |
| M80 | 45283 | 0.043 | -0.028 | 0.168 | 0.047 | 0.038 |
| M81 | 45283 | 0.040 | -0.101 | 0.238 | 0.037 | 0.065 |
| F1 | 45283 | 0.025 | -0.091 | 0.140 | 0.024 | 0.033 |
| F2 | 45283 | 0.072 | -0.297 | 0.297 | 0.077 | 0.089 |
| F3 | 45283 | 0.076 | -0.325 | 0.334 | 0.082 | 0.097 |
| F4 | 45283 | 0.230 | -0.030 | 0.683 | 0.204 | 0.143 |
| F5 | 45283 | 0.076 | -0.375 | 0.630 | 0.056 | 0.128 |
| F6 | 45283 | 0.101 | -0.389 | 0.769 | 0.074 | 0.156 |
| F7 | 45283 | 0.147 | -0.240 | 0.878 | 0.110 | 0.168 |
| F8 | 45283 | 0.081 | -0.366 | 0.675 | 0.059 | 0.133 |
| F9 | 45283 | 0.055 | -0.351 | 0.555 | 0.037 | 0.117 |
| F10 | 45283 | 2.053 | -53.255 | 71.056 | 0.437 | 13.216 |
| F11 | 45283 | 0.646 | 0.246 | 0.877 | 0.664 | 0.137 |
| F12 | 45283 | 0.207 | 0.002 | 0.518 | 0.185 | 0.118 |
| F13 | 45283 | 2.295 | 0.320 | 7.091 | 1.978 | 1.380 |
| F14 | 45283 | 0.703 | 0.008 | 2.747 | 0.580 | 0.526 |
| F15 | 45283 | 271024.630 | 2294.568 | 2059814.000 | 82875.887 | 440538.170 |
| F16 | 45283 | 257604.930 | 3862.360 | 1902089.000 | 82955.068 | 402718.020 |
| F17 | 45283 | 75891.037 | 178.341 | 1113313.100 | 15780.053 | 166830.080 |
| F18 | 45283 | 73339.253 | 1014.307 | 925537.000 | 27794.559 | 111710.830 |
| F19 | 45283 | 193532.010 | 891.086 | 1281551.900 | 56344.521 | 320996.480 |
| F20 | 45283 | 1.137 | -83.796 | 100.055 | 2.196 | 17.240 |
| F21 | 45283 | 2.937 | 1.334 | 10.551 | 2.405 | 1.608 |
| F22 | 45283 | 1.275 | 0.221 | 4.145 | 1.168 | 0.645 |
| F23 | 45283 | 0.804 | 0.159 | 3.234 | 0.712 | 0.486 |
| F24 | 45283 | 0.271 | 0.025 | 1.662 | 0.202 | 0.250 |
| F25 | 45283 | 0.100 | -0.339 | 0.552 | 0.075 | 0.188 |
| F26 | 45283 | 0.024 | 0.000 | 0.152 | 0.013 | 0.029 |
| F27 | 45283 | 0.323 | 0.101 | 0.727 | 0.304 | 0.122 |
| F28 | 45283 | 1.594 | 0.093 | 3.824 | 1.567 | 0.786 |
| F29 | 45283 | 1.025 | 0.888 | 1.294 | 1.019 | 0.054 |
| F30 | 45283 | 12.353 | 0.033 | 364.165 | 3.100 | 39.573 |
| F31 | 45283 | -0.017 | -0.195 | 0.151 | -0.016 | 0.053 |
| F32 | 45283 | 0.161 | 0.016 | 0.710 | 0.131 | 0.119 |
| F33 | 45283 | 0.009 | 0.000 | 0.124 | 0.003 | 0.018 |
| F34 | 45283 | 1818.989 | -46781.000 | 67511.000 | 602.409 | 12828.860 |
| F35 | 45283 | -2.225 | -15.501 | -0.254 | -1.519 | 2.214 |
| F36 | 45283 | 320.715 | -7014.682 | 9402.646 | 68.916 | 2056.968 |
| F37 | 45283 | 25729.760 | 70.946 | 197485.060 | 8236.659 | 40510.313 |
| F38 | 45283 | 44.737 | 0.953 | 963.420 | 9.417 | 105.688 |
| F39 | 45283 | 60.128 | 0.379 | 382.832 | 38.759 | 74.258 |
| F40 | 45283 | 15.106 | 0.115 | 484.529 | 4.460 | 49.558 |
| F41 | 45283 | 446.002 | 0.753 | 3155.774 | 81.831 | 696.922 |
| F42 | 45283 | 7.951 | -151.149 | 276.714 | 3.899 | 47.582 |
| F43 | 45283 | 12.353 | 0.033 | 364.165 | 3.100 | 39.573 |
| F44 | 45283 | 26.509 | 2.848 | 401.579 | 11.340 | 51.347 |
| F45 | 45283 | 1.671 | 0.072 | 16.535 | 0.860 | 2.472 |
| F46 | 45283 | 16.240 | 0.673 | 309.094 | 6.804 | 34.884 |
| F47 | 45283 | 0.953 | 0.302 | 1.281 | 1.011 | 0.221 |
| NF1 | 45283 | 6511.366 | 308.000 | 45585.033 | 2988.930 | 7899.995 |
| NF2 | 45283 | 276.987 | 75.000 | 472.000 | 284.000 | 90.774 |
| NF3 | 45283 | 76.477 | 62.780 | 87.050 | 76.780 | 5.211 |
| NF4 | 45283 | 67.119 | 48.410 | 87.390 | 66.550 | 8.353 |
| NF5 | 45283 | 79.629 | 53.930 | 100.000 | 80.000 | 8.323 |
| NF6 | 45283 | 79.444 | 57.950 | 90.670 | 80.470 | 6.072 |
| NF7 | 45283 | 0.706 | 0.000 | 1.000 | 1.000 | 0.455 |
| NF8 | 45283 | 1.508 | 1.000 | 3.000 | 1.000 | 0.605 |
| NF9 | 45283 | 8.309 | 0.000 | 80.990 | 0.000 | 16.951 |
| NF10 | 45283 | 4.882 | 0.000 | 17.000 | 4.000 | 3.861 |
| NF11 | 45283 | 22.120 | 0.000 | 356.000 | 4.000 | 55.209 |
| NF12 | 45283 | 0.443 | 0.000 | 1.000 | 0.500 | 0.335 |
| NF13 | 45283 | 0.270 | 0.000 | 1.000 | 0.250 | 0.278 |
| NF14 | 45283 | 0.107 | 0.000 | 1.000 | 0.000 | 0.190 |
| NF15 | 45283 | 25.769 | 0.210 | 100.000 | 17.710 | 25.563 |
| NF16 | 45283 | 27.031 | 0.640 | 89.460 | 22.340 | 20.242 |
| NF17 | 45283 | -0.100 | -1.560 | 4.802 | -0.207 | 0.585 |
| NF18 | 45283 | 0.273 | 0.000 | 1.000 | 0.000 | 0.446 |
| NF19 | 45283 | 0.913 | 0.000 | 4.000 | 0.000 | 1.559 |
| NF20 | 45283 | 0.484 | 0.000 | 1.000 | 0.000 | 0.500 |
| NF21 | 45283 | 0.988 | 0.000 | 1.000 | 1.000 | 0.109 |
| NF22 | 45283 | 1.837 | 1.000 | 2.000 | 2.000 | 0.370 |
| NF23 | 45283 | 0.378 | 0.083 | 0.802 | 0.377 | 0.158 |
| NF24 | 45283 | 1.643 | 0.000 | 37.533 | 0.006 | 5.386 |
| NF25 | 45283 | 3.168 | 0.000 | 46.799 | 0.005 | 8.591 |
| NF26 | 45283 | 11.127 | 5.000 | 21.000 | 11.000 | 3.357 |
| NF27 | 45283 | 5.169 | 3.000 | 13.000 | 5.000 | 2.210 |
| NF28 | 45283 | 37.731 | 12.500 | 66.667 | 36.364 | 10.343 |
| NF29 | 45283 | 6.103 | 0.600 | 25.970 | 3.497 | 6.035 |
| NF30 | 45283 | 0.215 | 0.000 | 1.000 | 0.000 | 0.411 |
| B1 | 45283 | 48.457 | 12.000 | 120.000 | 36.000 | 20.662 |
| B2 | 45283 | 24.992 | 1.000 | 105.000 | 22.000 | 18.044 |
| B3 | 45283 | 0.954 | 0.000 | 1.000 | 1.000 | 0.209 |
| B4 | 45283 | 2.242 | 0.000 | 4.477 | 2.303 | 0.768 |
| B5 | 45283 | 0.133 | 0.000 | 1.000 | 0.000 | 0.340 |
| B6 | 45283 | 0.445 | 0.000 | 1.000 | 0.000 | 0.497 |
| B7 | 45283 | 0.012 | 0.000 | 1.000 | 0.000 | 0.111 |
| B8 | 45283 | 0.307 | 0.000 | 1.000 | 0.000 | 0.461 |
| B9 | 45283 | 0.235 | 0.000 | 1.000 | 0.000 | 0.424 |

**Note:** In this table, some indicators with large absolute values (M20, F15-F19, F34, F36, F37, NF1,

NF29) have been scaled by 1,000,000.